\documentclass[twocolumns,showpacs,aps,floatfix,floats,endfloats]{revtex4}

\usepackage{graphicx} 
\usepackage{dcolumn} 
\usepackage{epsfig}
\usepackage{amssymb} 
\usepackage{amsmath} 
\usepackage{rotating} 
\usepackage{color}

\begin{document}

\title{Dynamical behavior of a complex fluid near an out-of-equilibrium transition:
approaching simple rheological chaos}
\author{Jean-Baptiste Salmon}
\affiliation{Centre de Recherche Paul Pascal, Avenue Schweitzer,
33600 PESSAC, FRANCE }
\author{Annie Colin}
\affiliation{Centre de Recherche Paul Pascal, Avenue Schweitzer,
33600 PESSAC, FRANCE }
\author{Didier Roux}
\affiliation{Centre de Recherche Paul Pascal, Avenue Schweitzer,
33600 PESSAC, FRANCE }

\date{\today}
\begin{abstract}
We report here an extensive study of sustained oscillations 
of the viscosity of a complex fluid
near an out-of-equilibrium transition. Using well defined
 protocols, we perform rheological 
measurements of the onion texture near a layering transition
 in a Couette flow. This complex fluid
exhibits sustained oscillations  of the viscosity, on a 
large time scale (500~s) at controlled stress.
These oscillations are directly correlated to an oscillating 
microstrutural change of the texture 
of the fluid.
We observe a great diversity of dynamical behavior and 
we show that there is a coupling with 
spatial effects in the $\nabla\!v$ direction. 
This is in agreement with a carefull analysis of
the temporal series of the viscosity with the dynamical
 system theory. This analysis indicates
that the observed dynamical responses do not strictly 
correspond to 3-dimensional chaotic states,
probably because some spatio-temporal effects are present and are likely to
play an important role.   

\end{abstract}
\maketitle

\section{Introduction}

Foams, pastes, liquid crystals, polymers or emulsions share all a common property~: 
when submitted to a shear flow, they exhibit different unusual behavior.
They are known as \textit{complex fluids}. Rheological properties of these types of materials 
have been extensively studied mainly because they exhibit non-Newtonian behavior. 
If in terms of mechanical behavior, these systems have been relatively well described, 
the question of the microscopic origin of these 
effects has only been addressed more recently.
Indeed, all complex fluids are composed of a macromolecular architecture
which leads to a coupling between their structures and the flow. 
In the 80's the development of techniques allowing to measure the structure of 
the fluids under shear allowed the community of physicists to have in parallel 
informations on the rheological behavior and on the microstructure 
under flow \cite{Clark:80,Safinya:90,Diat:93}. 
One of the questions addressed by this type of approach is related to the very nature of the coupling 
between structure and flow; indeed this coupling may modify 
the structure of the fluid undergoing shear flow. 
It is now quite clear that a pure mechanical approach is not 
enough to understand the experimental behavior \cite{Cambridge:97,Edimbourg:00}.
Recently, a lot of different experimental systems have shown that 
shear can be coupled to a thermodynamical phase transition 
\cite{Schmitt:94,Berret:97,Diat:95_2}
but new types of organisation may appear under shear which are not existing at rest 
\cite{Pujolle:01,Diat:93,Panizza:95}. Theoretical approaches tried to described 
this complex behavior as a coupling between hydrodynamics and thermodynamics. 
Even though some success in this way can be noticed 
\cite{Onuki:87,Cates:89,Bruinsma:91,Olmsted:92,Ajdari:98,Sollich:97}
we are far from having a satisfying description of the steady 
behavior of these systems under shear.

Besides the understanding of the structure under steady 
shear and because these systems exhibit out-of-equilibrium 
transitions we do expect an even richer behavior. 
Indeed, it has been shown experimentally, 
that near out-of-equilibrium transitions, 
the temporal behavior of the viscosity of a complex fluid, namely lyotropic systems, 
may exhibit sustained oscillations 
\cite{Hu:98,Meyer:99,Bandyopadhyay:00,Fisher:00,Wunenberger:01,Cristobal:these}. 
The origin of the latter is still unknown, but some authors suggest different scenarios~: coupling
with shear induced structures \cite{Wunenberger:01}, mechanical instability in a shear banding
case \cite{Bandyopadhyay:00} or coupling with elastic instabilities \cite{Fisher:00}. 
It has been also suggested that the rheology of wormlike micelles can exhibit 
chaotic behavior \cite{Bandyopadhyay:00}. Theoretically, with a microscopic model
some authors found \textit{rheo-chaos} in the rheology of a nematic liquid crystal 
\cite{Grosso:01}.
Other authors proposed  
theoretical models \cite{Head:01}, based on the equations of the rheology of 
soft glassy materials \cite{Sollich:97}, in which sustained oscillations 
of the shear rate at an imposed stress take place. Very recently, the same authors
have found \textit{rheo-chaos} in such spatially homogenous models \cite{Cates:02}.
The aim of this paper is to make an extensive study of an experimental 
system where sustained oscillations have been previously observed \cite{Wunenberger:01}. 
In the system studied here, a close-compact assembly of soft 
elastic spheres (onions) \cite{Leng:01,Sierro:97},
it has been early assessed \cite{Roux:93} 
that the theory of bifurcations may be a guide for
the understanding of the temporal rheological behavior of the complex fluid. This article 
presents new experimental results concerning the great
diversity of temporal observed responses
near an out-of-equilibrium transition. In a first step
we carefully study using several protocols the temporal behavior
of the rheological signals. We try to evidence a dynamical scenario 
which may indicate the presence of an Hopf bifurcation. 
In a second part we relate this rheological behavior to the 
structural evolution of the fluid using light scattering. The last part 
of the work is related to a carefull analysis of the complex 
temporal behavior we get in certain cases (resembling to chaos). We show 
with the help of dynamical system theory that 
the oscillating viscosity may not simply be
described with a 3-dimensional dynamical system, probably because 
spatio-temporal effects are playing an important role. 

\section{Dynamical behavior of the rheology of the onion texture}

The complex fluid we consider here is a lyotropic lamellar phase prepared with SDS (6.5~\%wt),
octanol (7.8~\%wt)  and water salted with 20~g.l$^{-1}$ of sodium chloride.
At equilibrium, this phase is made of membranes (surfactant bilayers) 
of thickness $\delta= 2$~nm,
regularly stacked and separated by solvent.
The distance between the lamellae, the smectic period d, measured with neutron scattering,
is about $15$~nm \cite{Herve:93}. Such a system is stabilized by entropic
interactions due to exclusion between the undulating membranes \cite{Helfrich:78}.
This system is sensitive to temperature : at $34^{\circ}$C, a 
diphasic sponge-lamellar mixture appears \cite{Porte:91,Roux:92}.

\begin{figure}[ht]
\begin{center}
\epsfxsize= 8.0cm
\centerline{\epsfbox{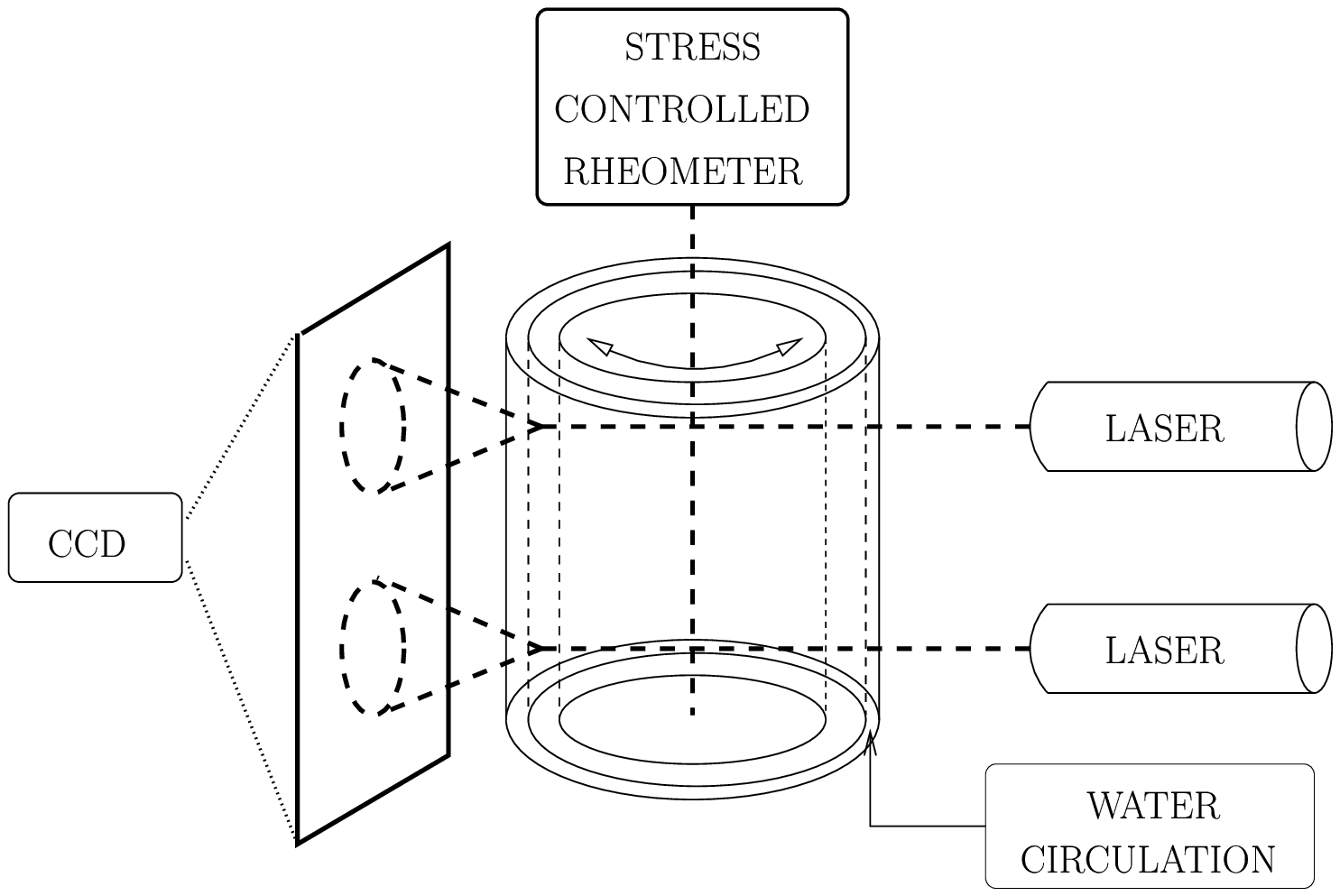}}
\end{center}
\caption{Experimental setup, a thermostated plate (not shown) on the top of the cell 
allows to avoid evaporation.}
\label{setup}
\end{figure}

To probe the rheological properties of this complex fluid, we use the 
experimental device presented in Fig.~\ref{setup}.

In order to measure the viscosity, the lamellar phase is confined 
in Mooney-Couette cells with gap $e = 1$~mm or $e = 0.5$~mm,
height $h = 30$~mm and inner radius $R = 24$~mm. A stress controlled rheometer 
(AR1000 TA Instrument) allows us to impose
a torque on the axis, on which the rotor is fixed, and so to induce a 
controlled stress $\sigma$ in the fluid.
The rheometer records continuously the shear rate $\dot{\gamma}$ as a function of time, the time scale
of this measurement is very short compared to the time scales of the experiment.
In order to follow the effect of shear on the microstructure 
of the phase, the cell is totally
transparent and two lasers (He-Ne 15mW) give the diffraction patterns
at different heights in the cell. The patterns are collected on a screen and digitalized 
with a CCD camera (Cohu).
Since the laser beams go trough the sample twice,
one obtains two diffraction rings. The diffraction pattern corresponding to the first sample
has an ellipsoidal shape due to the optical deformation of the Couette cell (playing
the role of a cylindrical lens). The second sample leads to a classical circular ring.
Temperature $T$ is controlled within $\pm~0.1^{\circ}$C using a water circulation around the cell 
(the range of variation of the latter is about $\pm~0.04^{\circ}$C).
The experimental behavior observed here, depends strongly on the sample, 
namely on the concentration of octanol.
With a classical setup, we observed that after a few hours (2-3~hours), 
we have a significant change 
in the composition due to the evaporation of octanol and water.  
To control the evaporation of octanol and water and their condensation on the top of the cell,
the latter is closed with a thermostated plate, which allows up
to 80-hours experiments with the same sample and a negligible evaporation.

\begin{figure}[ht]
\begin{center}
\epsfxsize= 3.0cm
\centerline{\epsfbox{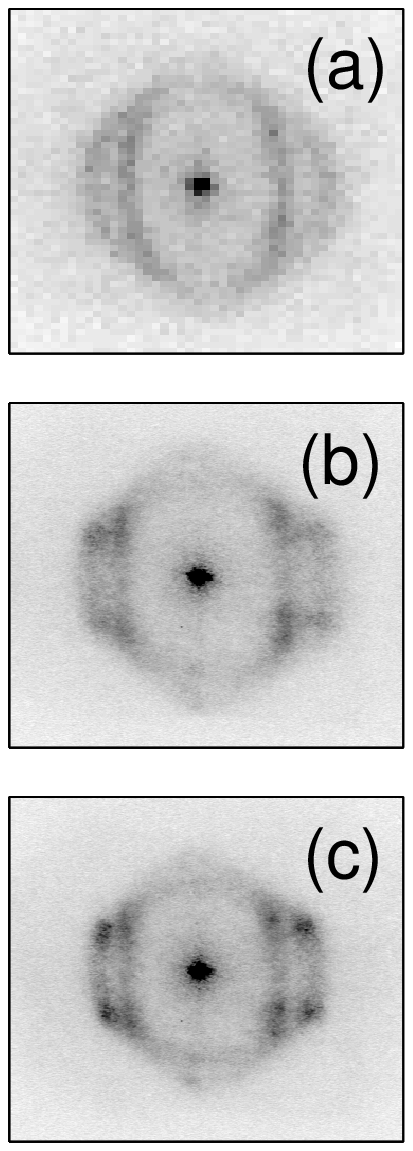}}
\end{center}	
\caption{Diffraction figures obtained under shear. (a) Ring of scattering, (b) Fuzzy hexagonal 
pattern, (c) Hexagonal pattern.}
\label{diffraction}
\end{figure}

The effect of shear on this system has been extensively studied previously 
\cite{Sierro:97,Leng:01}.
It has been shown that shear controls the texture
of the lamellar phase and
series of textural transitions are observed as $\dot{\gamma}$ is increased.
Small angle light scattering allows us to characterize the different textures which 
can be obtained at different stress. 
At low shear rates, a state of partially oriented lamellar phase is observed; at a typical shear
rate of $1$~s$^{-1}$, the texture changes radically~: the membranes are wrapped
in multilamellar vesicles (called onions), close compact organized. The size of
these onions is of the order of microns, and scales with shear rate according
to $ R \sim \dot{\gamma}^{-\frac{1}{2}}$. The diffraction
pattern is an homogeneous ring (cf. Fig.~\ref{diffraction}(a)), indicating that
there is no long range correlations between onions.

At $\dot{\gamma} \approx 15~$s$^{-1}$, six fuzzy peaks 
appear on the ring (cf. Fig.~\ref{diffraction}(b)).
This corresponds to the \textit{layering} transition~\cite{Sierro:97,Diat:95}~:
the onions exhibit now a long-range orientational order under shear (which is 
conserved and evolved into a long-range positional order when the shear is stopped) 
\cite{Sierro:97,Leng:01}.
In this regime, two dimensional
layers of onions with hexagonal order, slide on each other.
This transition is named the layering transition after a similar disorder/order 
transition was observed in colloids under shear \cite{Ackerson:84}. 
When the shear rate is increased, the peaks on the ring become more contrasted
as shown in Fig.~\ref{diffraction}(c).
It is impossible to assess whether the diffraction pattern shown in Fig.~\ref{diffraction}(b)
corresponds to a coexistence between the
two different textures or to weak spatial correlations. 
The spots with wave vectors along the rheometer axis are less 
intense than the others, this is
due to the zig-zag motion of the plans of onions when sliding on each other \cite{Ackerson:84}.

\begin{figure}[ht]
\begin{center}
\epsfxsize= 8.0cm
\centerline{\epsfbox{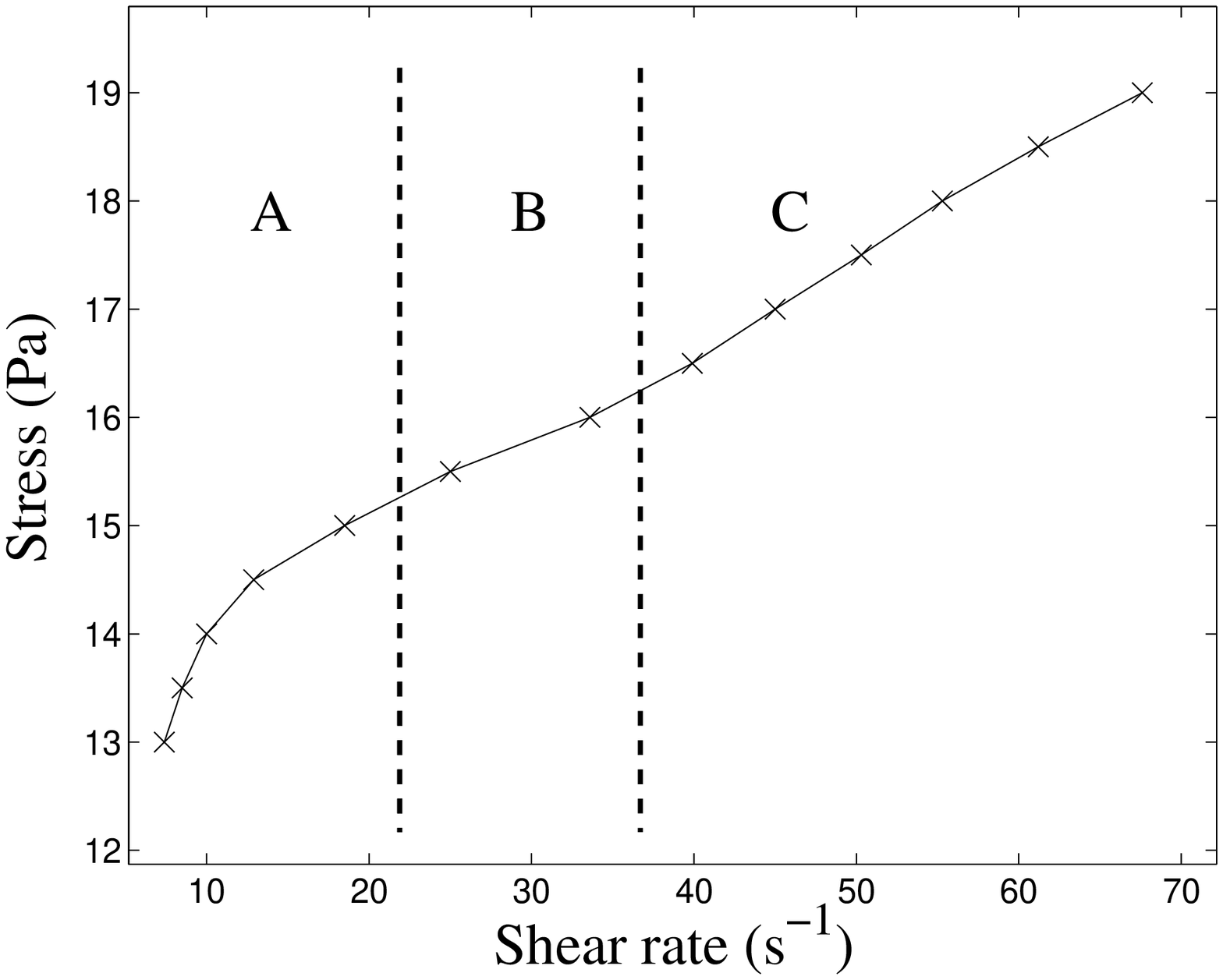}}
\end{center}	
\caption{Flow curve at $T=26^\circ$C, $e=1$~mm. The regions A, B and C correspond 
to the different diffraction patterns shown in Fig.~\ref{diffraction}.}
\label{flowcurveT26}
\end{figure}

Actually the layering transition exhibits different 
rheological behavior when temperature is changed.  
When $T \leq 27^{\circ}$C,
the rheological flow curves $\sigma$ vs $\dot{\gamma}$ are continuous.
It is always
possible to define an asymptotic stationary
value for the measured shear rate. A typical flow curve $\sigma$ vs $\dot{\gamma}$, 
for $T=26^{\circ}$C
and $e=1$ mm is shown in Fig.~\ref{flowcurveT26},

\begin{figure}[ht]
\begin{center}
\epsfxsize= 8.0cm
\centerline{\epsfbox{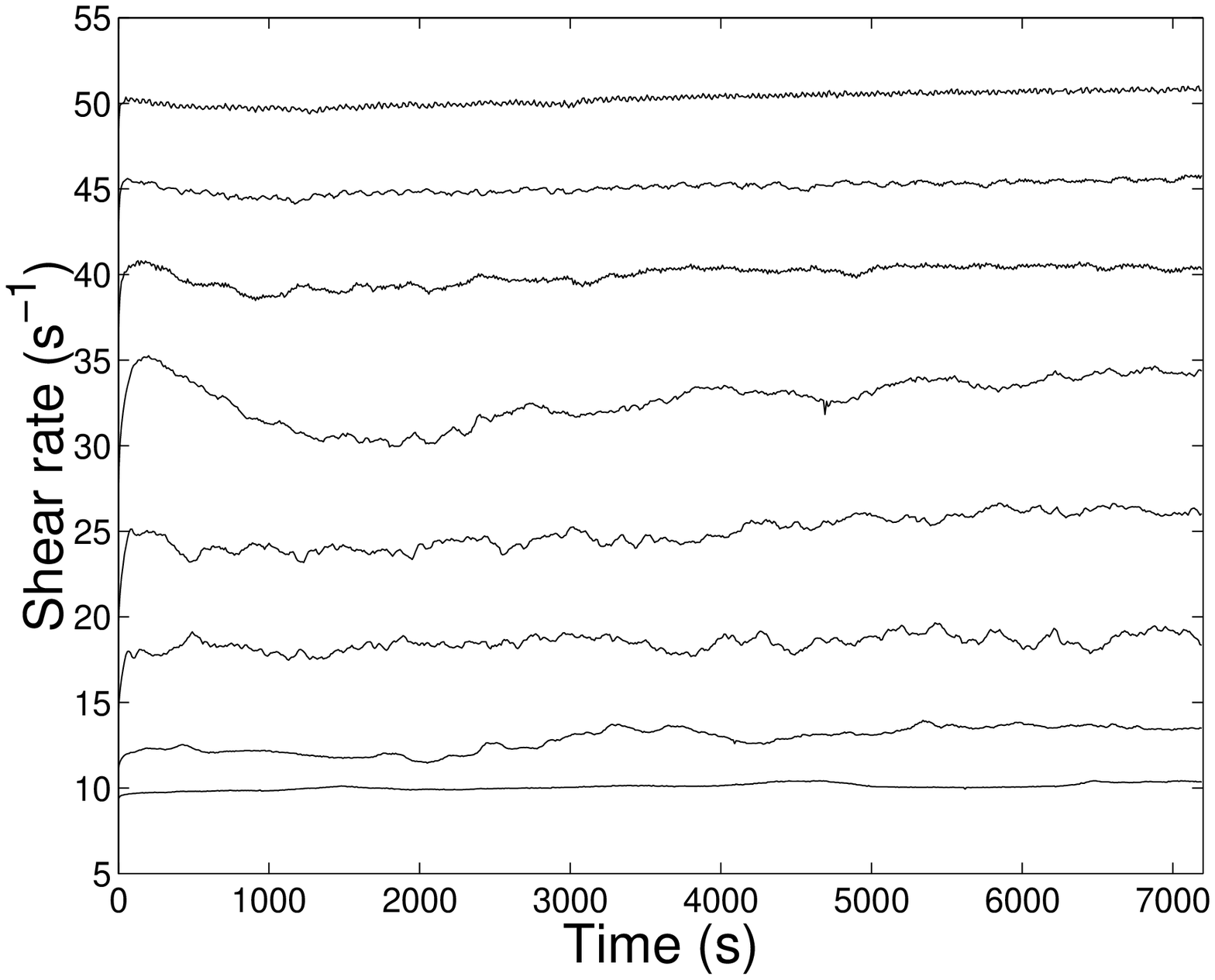}}
\end{center}	
\caption{Temporal responses of the shear rate for the range $14 \rightarrow 17.5$ Pa 
of imposed stress with increment $\delta\sigma = 0.5$~Pa, 
corresponding to the flow curve shown in Fig.~\ref{flowcurveT26}.}
\label{amplitudeT26}
\end{figure}

each point corresponds to a stationary state of the shear rate.
The different regions A, B and C shown in the flow curve correspond
to the different diffraction patterns
respectively shown in Fig.~\ref{diffraction}(a), (b) and (c).
In Fig.~\ref{amplitudeT26} are shown the temporal responses of the shear rate for 
the range $14 \rightarrow 17.5$~Pa and increment of stress of 0.5~Pa.
The temporal behavior of the shear rate
become noisier as the critical stress $\sigma_c \approx 16$~Pa is
approached. The Fourier transforms of these time series exibit no peaks so 
there is no characteristic time in the recorded noisy shear rate.

\begin{figure}[ht]
\begin{center}
\epsfxsize= 8.0cm
\centerline{\epsfbox{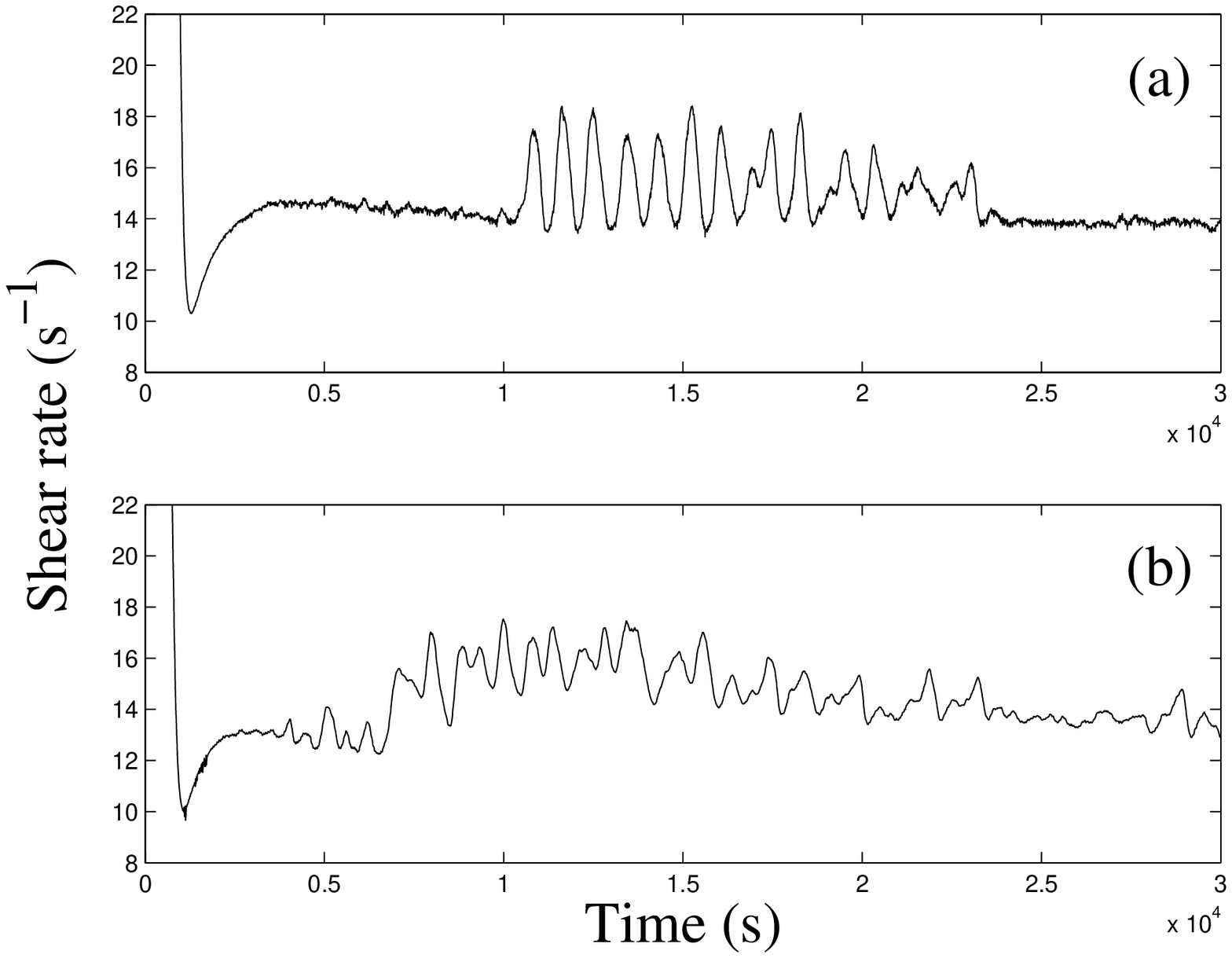}}
\end{center}	
\caption{Typical shear rate responses at an imposed stress ($\sigma=15$ Pa, $T=31^{\circ}$C)  
near the layering transition.}
\label{longtime}
\end{figure}

When $T \geq 27^{\circ}$C and stress is imposed, there is a region where it is difficult
to define a stationary viscosity.  
Typical shear rate responses, on a fresh sample, at an imposed stress near the critical
stress $\sigma_c$ of the layering transition and at a temperature $T = 31^{\circ}C$
are shown in Fig.~\ref{longtime};
 the stationary disordered onion texture appears after
about 30~minutes as seen with light scattering. In Fig~\ref{longtime}(a), after 10000~s, 
the recorded shear rate exhibits
transitory oscillations with a period of about 500~s, this oscillating regime
disapears after 6/7 hours and an 
\textit{asymptotic stationary} state
is reached after 25000~s : the nonlinear rheology of this system exhibits
very long time of asymptotisation
near the out-of-equilibrium transition. The second recorded behavior reproduced 
in Fig.~\ref{longtime}(b) 
has been obtained with the same stress. The asymptotic state for the shear rate
differs from the first experiment, this state corresponds 
to a noisy complex dynamical response~: no stationary shear rate 
can be easiliy defined. 
This illustrates the strong dependence on initial conditions for the onset of the temporal 
instabilities.  

\begin{figure}[ht]
\begin{center}
\epsfxsize= 8.0cm
\centerline{\epsfbox{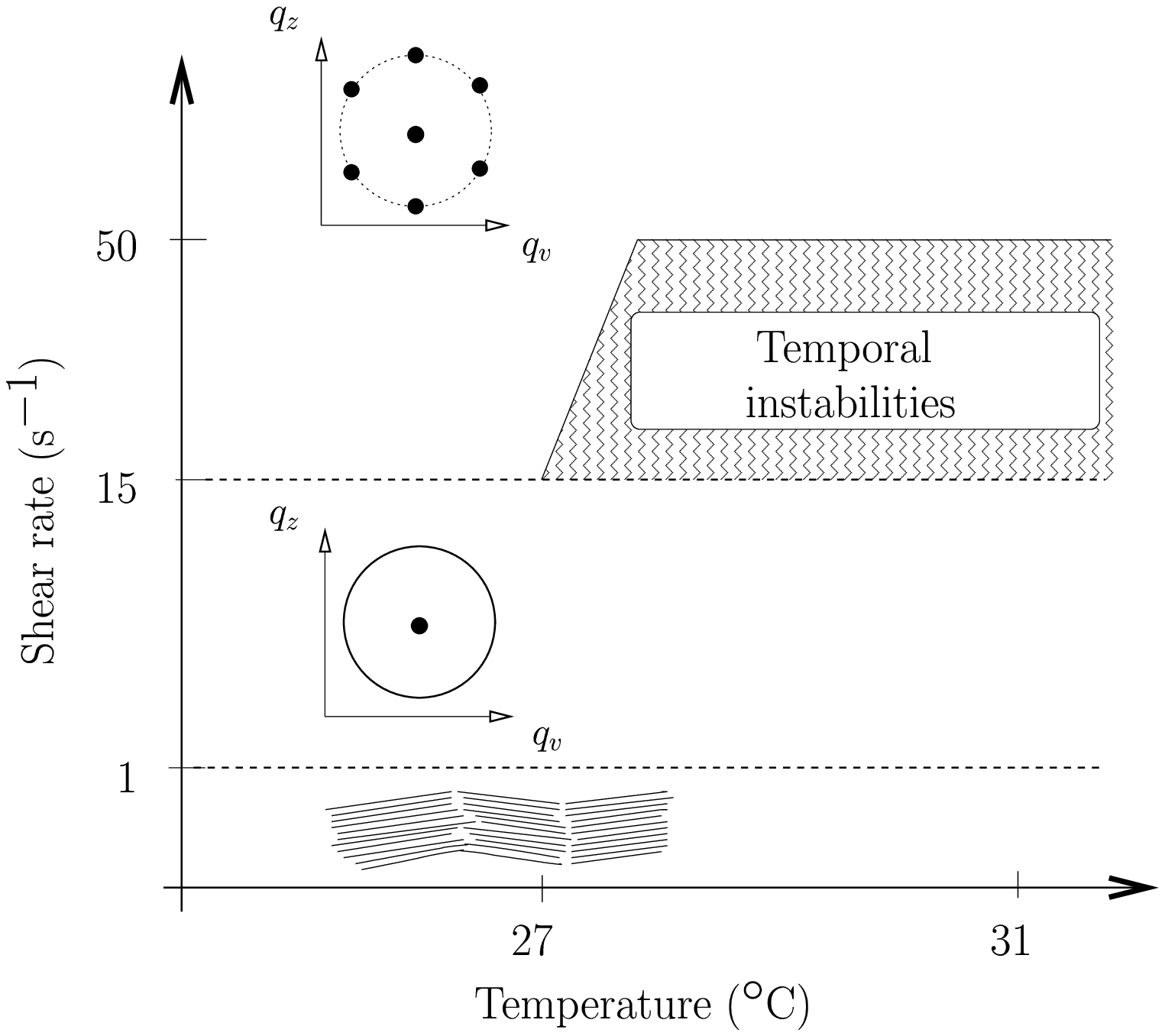}}
\end{center}
\caption{Shear diagram of the lyotropic lamellar phase studied, the gray region 
corresponds to non-stationary shear rate responses at stress controlled.}
\label{orientation}
\end{figure}

The region of dynamical behavior of the 
viscosity seems to vanish when $T$ approaches $27^\circ$C. A convenient way to represent 
all these effects is to use a \textit{shear diagram}, where \textit{stationary}
textures are plotted as a function of $T$ and $\dot{\gamma}$. Such a diagram is plotted
in Fig.~\ref{orientation}.  

As suggested in a precedent work \cite{Wunenberger:01}, in the 
region of parameters where dynamical reponses occur, 
the rheological behavior 
represented by the flow curve must be  associated
with the temporal responses of the shear rate.    
In order to show the different \textit{asymptotic} dynamical responses of the shear rate
in the vinicity of the layering transition, we must define a protocol in order to
get enough reproducibility in the flow curves $\sigma$ vs $\dot{\gamma}$. Two parameters
are important~: the stress increment $\delta\sigma$ between two different imposed stress,
and the time interval $\delta t$ we wait before changing the applied stress. If
$\delta t \leq 1000$~s the different flow curves are not reproducible
and depend drastically on the
initial conditions. If $\delta\sigma \geq 1$~Pa we may miss the dynamical
region because of its narrowness.
Protocols with large $\delta t$ and small $\delta\sigma$ will correspond to
\textit{quasi-static} approaches of the transition. Actually, compromises have been 
found to use the most quasi-static approach. We are limited by the evaporation
of the sample which gives us a maximum of the accessible experimental time (about 80 hours).

\begin{figure}[ht]
\begin{center}
\epsfxsize= 8.0cm
\centerline{\epsfbox{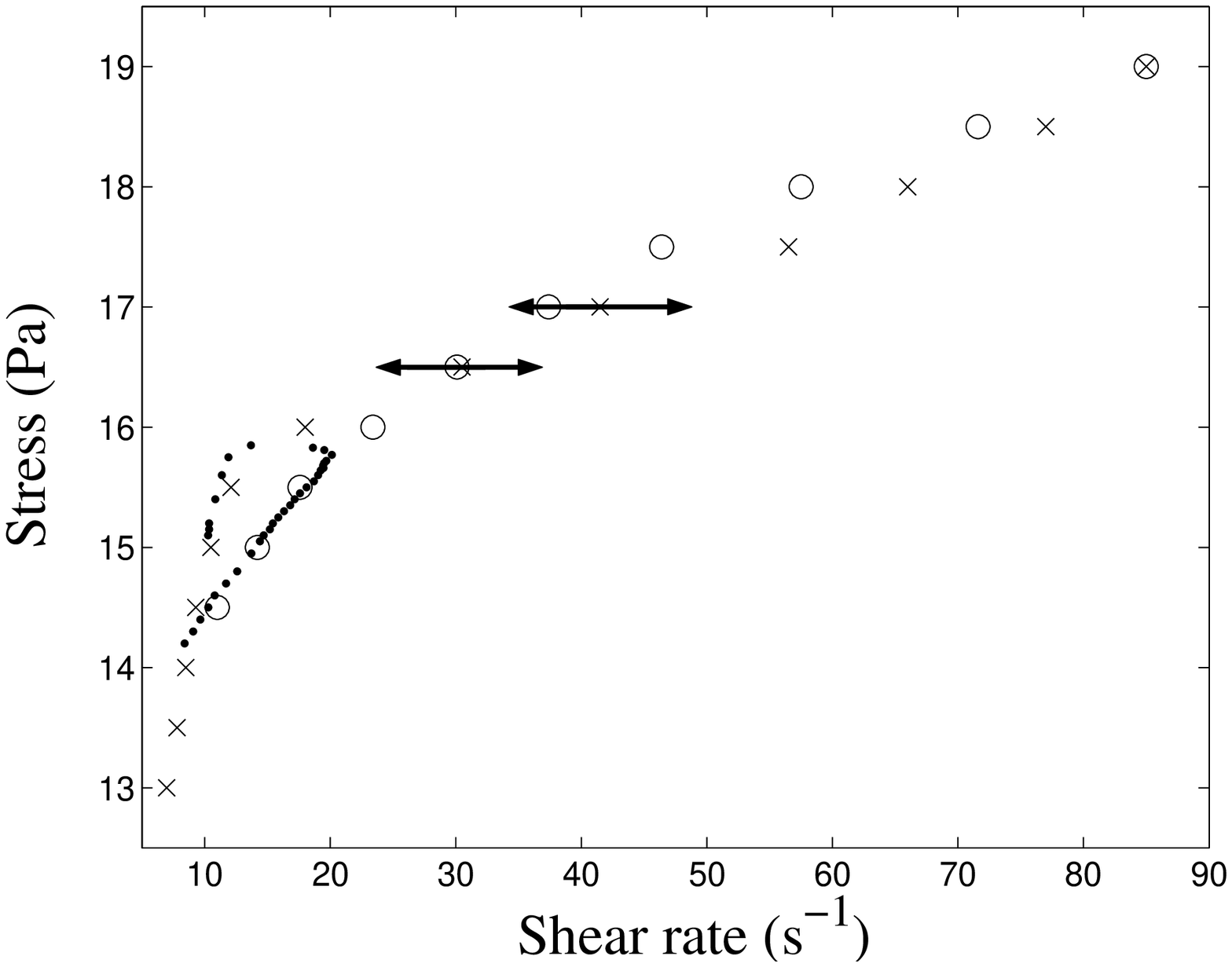}}
\end{center}
\caption{Different flow curves with $T=30^{\circ}$C and $e = 1$~mm, 
for two different experimental procedures. $\times$ correspond to protocol I, stress up. $\circ$
correspond to protocol I, stress down. $\cdot$ correspond to protocol II.}
\label{flowcurve1gap}
\end{figure}

\begin{figure}[ht]
\begin{center}
\epsfxsize= 8.0cm
\centerline{\epsfbox{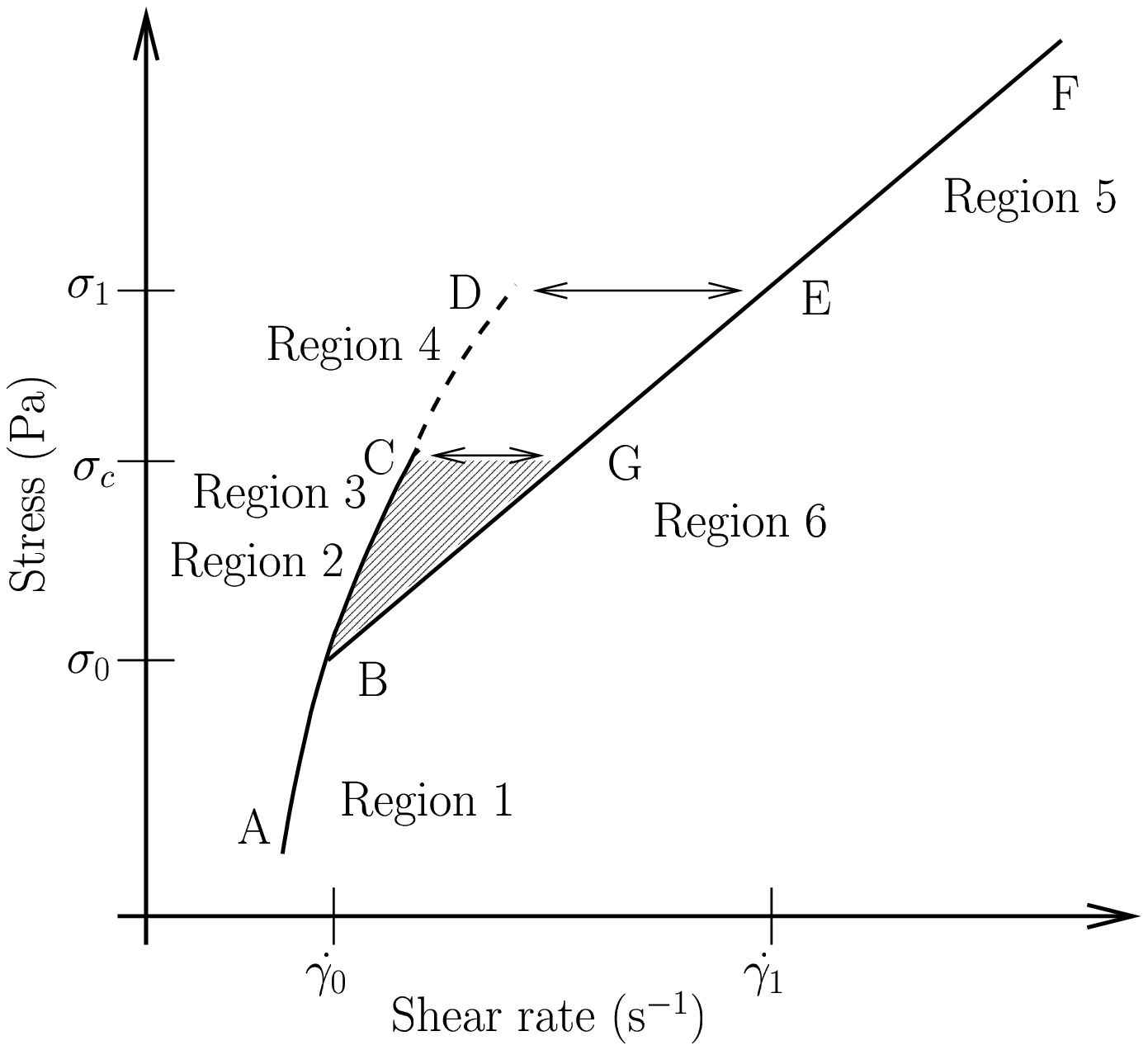}}
\end{center}
\caption{Schematic representation of the asymptotic states (solid lines), 
metastable states (dashed line)
in the vinicity of the layering transition, and their 
corresponding dynamical regions.} 
\label{scenario}
\end{figure}

For a systematic study we decided to use two different protocols to test 
the quality of the quasi-static approach we have. We have also used different 
geometries to try to separate temporal dependence instabilities from spatial structures 
(cf. Sec.~III). For that we made several Couette cells corresponding to different 
heights and different gaps. We will mainly discuss here the effect of the gap~:
two different ones have been studied (1~mm and 0.5~mm). 
Fig.~\ref{flowcurve1gap} shows two flow curves
for two different protocols, both of them with a gap $e=1$~mm.
In protocol I,
$\delta t = 7200$~s and $\delta\sigma = 0.5$~Pa,
the stress is first increased from 13 to 19~Pa and then decreased from 19 to 13~Pa.
The results of the second protocol (protocol II, $\delta t \approx 15000$~s 
and $\delta\sigma \approx 0.1$~Pa) will be discussed later.
The different values reported in Fig.~\ref{flowcurve1gap} correspond
to the mean values of the asymptotic recorded shear rate, and   
arrows represent oscillations with great 
amplitude between the maximum 
and the minimum value
of the oscillating shear rate.
Figure~\ref{scenario} shows schematically for more convenience, the different results
of these protocols. Continuous lines correspond to asymptotic stable states and dashed lines
to metastable states.

\begin{figure}[ht]
\begin{center}
\epsfxsize= 8.0cm
\centerline{\epsfbox{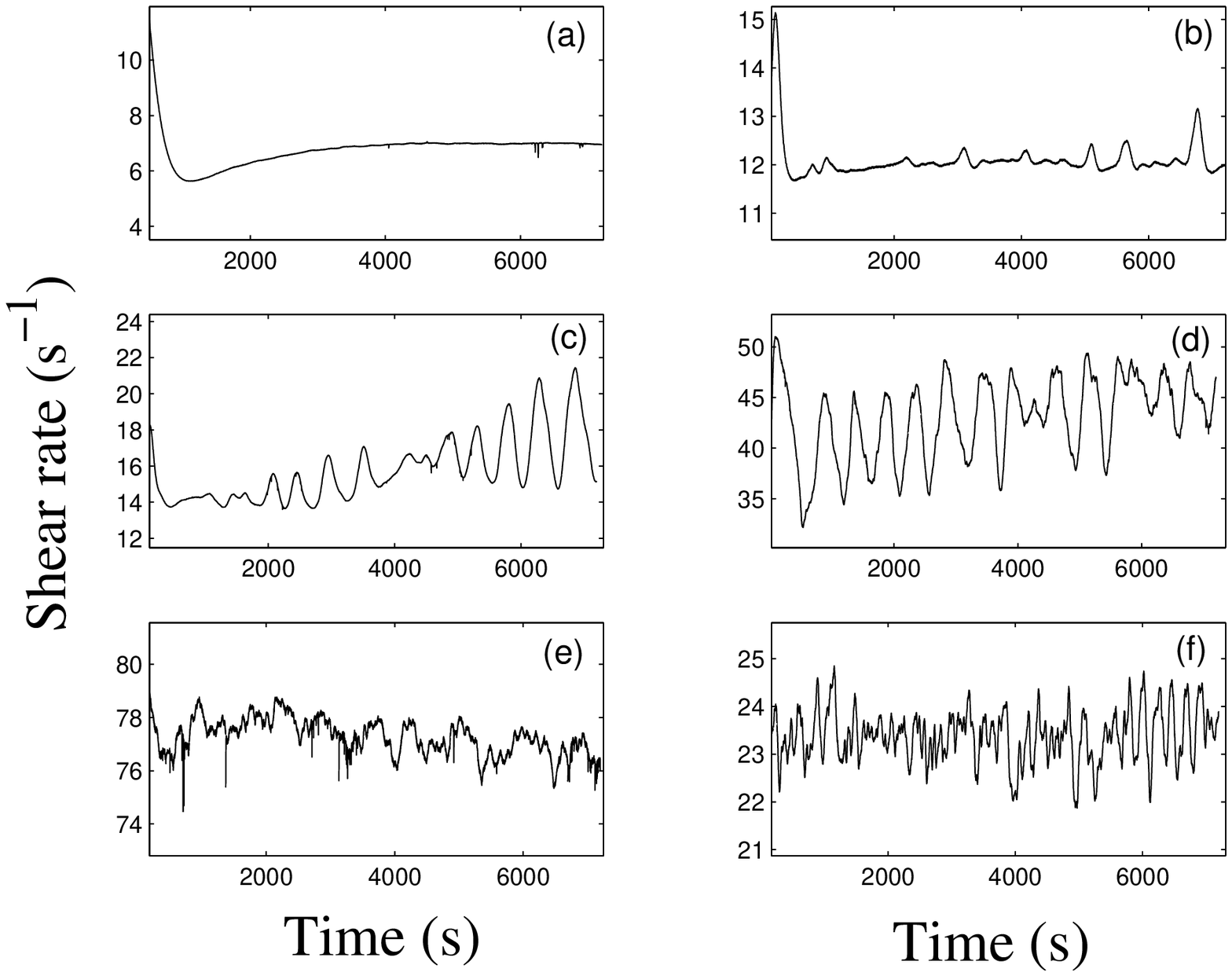}}
\end{center}
\caption{Different dynamical responses of the shear rate 
in the vinicity of the layering transition with $e=1$ mm. Note the different amplitudes
of these temporal reponses.
(a) Region 1 ($\sigma=13$Pa), 
(b) Region 2 ($\sigma=15.5$Pa), (c) Region 3 ($\sigma=16$Pa), (d) Region 4 ($\sigma=17$Pa),
(e) Region 5 ($\sigma=18.5$Pa), (f) Region 6 ($\sigma=16$Pa).}       
\label{sixregion}
\end{figure}

The flow curve, with protocol I, 
exhibits six regions of different temporal responses of $\dot{\gamma}$
displayed in Fig.~\ref{sixregion}. 

On the way up~: 
\begin{itemize}
\item
Region 1 ($A \rightarrow B$)~: relaxation to a stationary state of disordered onions 
(Fig.~\ref{sixregion}(a)).
\item 
Region 2 ($B \rightarrow C$)~: this is a branch which is followed on the way up,
noise appears with a characteristic period of 500 s and
an amplitude of about $1$~s$^{-1}$  (Fig.~\ref{sixregion}(b)).  
Note that even when we wait a very long time with a different protocol
no transition seems to appear to the branch $B \rightarrow G$ 
(we will see that this is different from the branch $C \rightarrow D$).

\item
Region 3 ($C$)~: the shear rate begins to oscillate after transient phase with a period of 500 s
and an amplitude of about $5$~s$^{-1}$ (Fig.~\ref{sixregion}(c)).  
At the same time six fuzzy peaks appear on the diffraction ring, indicating the onset of
spatial correlation between onions (cf. Fig~\ref{diffraction}(b)).
\item
Region 4 ($C \rightarrow D$)~:  the shear rate shows oscillations with a 
large amplitude of  $15~$s$^{-1}$ (this corresponds roughly to 
the distance between the branches $C \rightarrow D$ and $G \rightarrow E$) 
and with a period of about 500~s (Fig.~\ref{sixregion}(d)).   
In this regime the diffraction pattern
clearly shows a temporal correlation between the structure of the sample and
these oscillations. Modulation of the scattering pattern is observed on the time scale of
the rheological signal. When we wait enough time we end up going 
to the branch $G \rightarrow E$ which is the one followed the way down. 
\item
Region 5 ($E \rightarrow F$)~: the shear
	rate relaxes on a noisy stationary branch (Fig.~\ref{sixregion}(e)). The corresponding texture corresponds 
to the diffraction pattern shown in Fig~\ref{diffraction}(c). 
\end{itemize}

On the way down~: 
\begin{itemize}
\item
Region 6 ($E  \rightarrow B$)~: a complex dynamical state
appears progressively. The main period is 300~s and the greatest amplitude
is about $5$~s$^{-1}$ (Fig.~\ref{sixregion}(f)). 
This complex dynamics disappears also when
approching B.
\end{itemize}

\begin{figure}[ht]
\begin{center}
\epsfxsize= 8.0cm
\centerline{\epsfbox{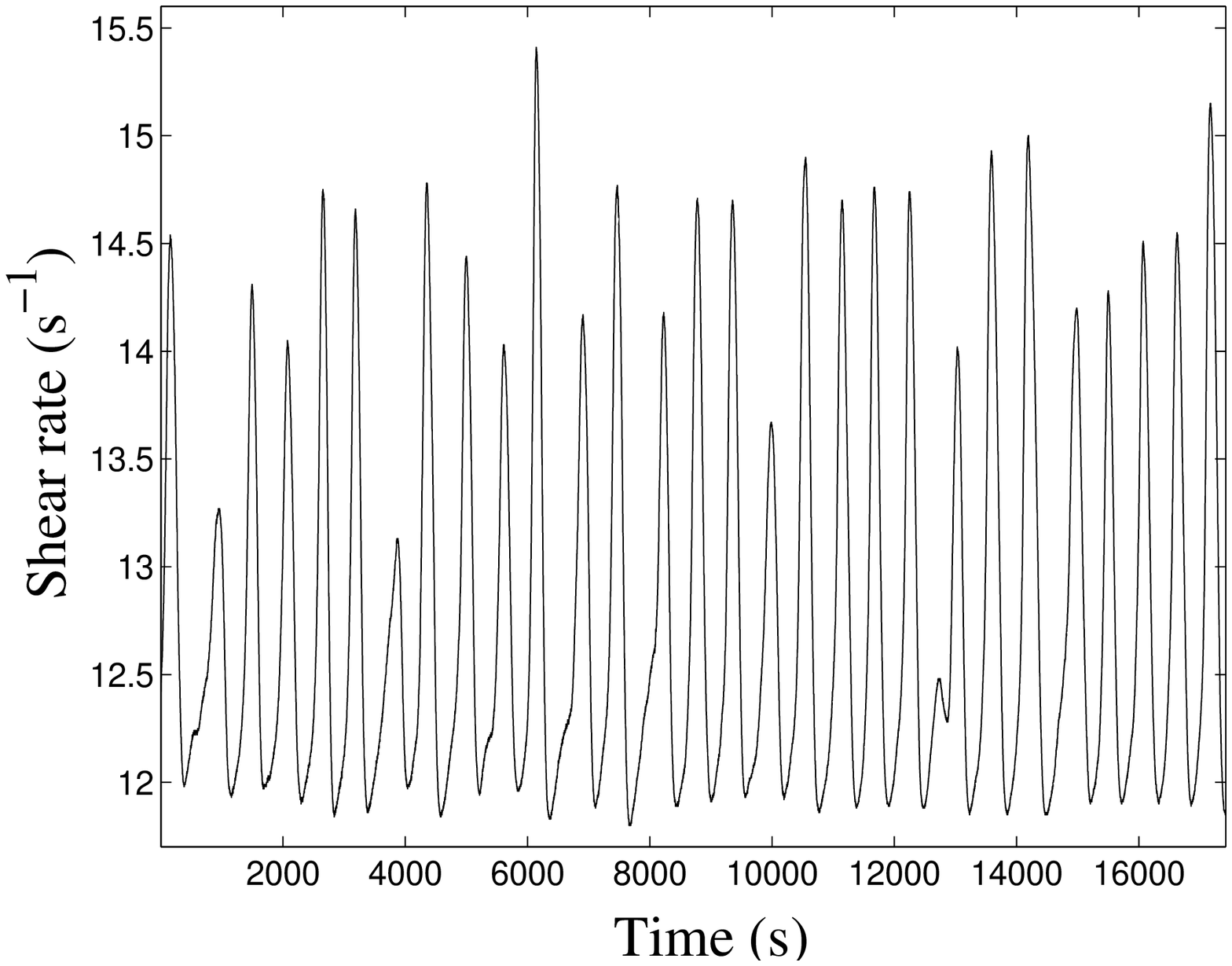}}
\end{center}
\caption{Aperiodic time serie $\dot{\gamma}(t)$ in region 3.}
\label{aperiodic}
\end{figure}

We will come back on the problem of analysing the signal in the last 
part of this article. Note however than on the way down the branch followed is different
from the one followed on the way up.
Protocol I allows to record transitory behavior which reveal hysteretis loop and 
oscillating viscosity. As predicted by the dynamical system theory, one can
expect richer dynamical behavior when asymptotic states are reached in the vinicity
of hysteretis and oscillating bifurcations \cite{Richetti:87,Argoul:87}.
We have seen that transient phases longer than $\delta t = 7200$~s may occur
with this system. So in order
to get a better understanding, the experiments were repeated with $\delta t \approx 15000$ s
and $\delta\sigma \approx 0.1$~Pa (protocol II on Fig.~\ref{flowcurve1gap}). In this protocol,
stress is imposed from B to C. At the point C, a slow drift of  
shear rate into the region 6 to the point G takes place.
Then the stress is imposed from G to A.
With the protocol II, region 4 does not exist, the way followed is region~2 $\rightarrow$
region~3 $\rightarrow$ region~6.   
The large oscillations of region 4 observed in protocol I correspond to metastable
dynamics (i.e they do not exist if we wait enough time).
This protocol allows to record asymptotic states of the oscillating viscosity~: in the vinicity
of the point C (region 3), aperiodic oscillations of the shear rate are recorded 
with a period of 500~s and an amplitude of $3$~s$^{-1}$ (Fig.~\ref{aperiodic}).

This aperiodic time serie corresponds to oscillations  
between a fixed minimum value ($\approx 12$~s$^{-1}$) and a non-fixed maximum value. 
The region of existence of the latter is very thin,
about 0.05~Pa and the transient phases take about 5 hours,
so up to 80 oscillations have been recorded before the slow drift 
($C \rightarrow G$) of the shear rate into the
region of complex dynamics (region 6) takes place.
The dynamics in region 6 shown in Fig.~\ref{region6} 
corresponds to a complex dynamical state, with a 
period of about about 300~s. 
When stress goes down approaching point B, the quasi-static protocol II allows to show the
simplification shown in Fig.~\ref{simplification} of the complex dynamics of region 6.
To the precision of the protocol ($\delta t \approx 15000$~s and  $\delta\sigma \approx 0.1$~Pa),
no simplier dynamics has been recorded. The main difference 
between the two protocols (i.e. the disparition of region 4) is a well known characteristic
of hysteretis behavior~: the hysteretis loop is larger 
($B \rightarrow D  \rightarrow E \rightarrow B$ 
in protocol I) with small intervall of time $\delta t$ than
for larger $\delta t$  ($B \rightarrow C  \rightarrow G \rightarrow B$ in protocol II) \cite{Grand:97}. 
With the protocol II we can assess that the loop ($B \rightarrow C  \rightarrow G \rightarrow B$)
corresponds to a real hysteresis loop between region 2 and region 6. 

\begin{figure}[ht]
\begin{center}
\epsfxsize= 8.0cm
\centerline{\epsfbox{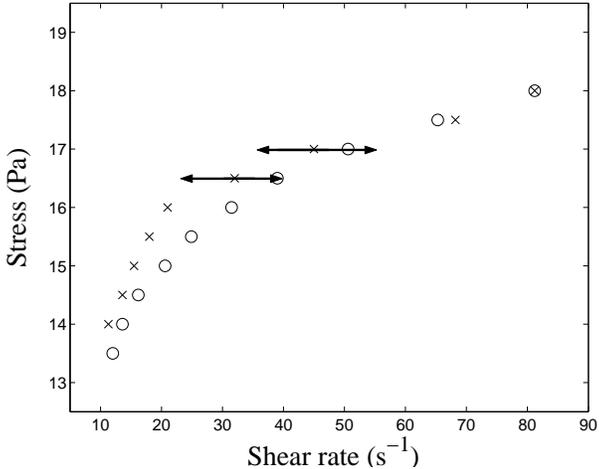}}
\end{center}
\caption{Flow curve with $T=30^{\circ}$C and $e = 0.5$~mm for the experimental protocol I.
 $\times$ correspond to protocol I, stress up. $\circ$
correspond to protocol I, stress down.}
\label{flowcurve05gap}
\end{figure}

\begin{figure}[ht]
\begin{center}
\epsfxsize= 8.0cm
\centerline{\epsfbox{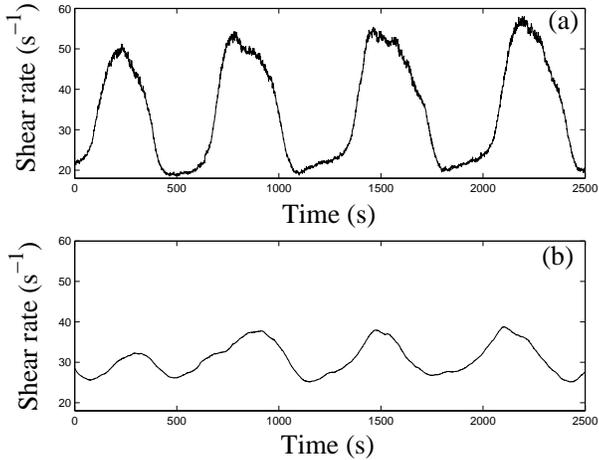}}
\end{center}
\caption{Shear rate oscillations in region 4 for two different geometries, (a) $e=0.5$~mm, 
(b) $e=1$~mm.} 
\label{relaxation}
\end{figure}

The same rheological measurements were performed with a gap 
$e=0.5$~mm, with the protocol I but from 14.5 to 18~Pa then from 18 to 13.5~Pa, 
the corresponding flow curve is shown in Fig.~\ref{flowcurve05gap}. 
The flow curves with the gaps $e=0.5$ and $e=1$~mm are 
similar. The confinement does not affect strongly the complexity we observed because
all the dynamical scenario (i.e. the six regions in a non-asymptotic protocol)  
is observed. The hysteresis loop and the different oscillating regimes still exist in the small gap.
However some rheological differences must be noticed~:
\begin{itemize}
\item
In a non-asymptotic protocol, as shown in fig.~\ref{relaxation}, 
larger oscillations in region 4 
(20~s$^{-1} \rightarrow$ 50~s$^{-1}$à) are obtained with $e=0.5$~mm
more easily than with the larger gap (25~s$^{-1} \rightarrow$ 40~s$^{-1}$). 
They have also a more relaxational shape
than for the larger gap (i.e. the transition from low values
to high values is sharper).
\item
The \textit{hysteresis} cycle ($B \rightarrow D \rightarrow E \rightarrow B$)
still exists, but is larger in the confined geometry
than for $e=1$~mm (cf. Figs.~\ref{flowcurve1gap} and \ref{flowcurve05gap},
$\dot{\gamma}_0 \approx 15$~s$^{-1}$, $\dot{\gamma}_1 \approx 55$~s$^{-1}$ for $e=0.5$~mm
and $\dot{\gamma}_0 \approx 15$~s$^{-1}$, $\dot{\gamma}_1 \approx 45$~s$^{-1}$ for $e=1$~mm).  

\item
With the gap $e=1$~mm, the branch $F \rightarrow G$ does not strictly correspond
to the branch $G \rightarrow F$ (cf. Fig.~\ref{flowcurve1gap}).   
\end{itemize}
These effects will be discussed in the next Section.
We have seen that a very rich temporal dependence behavior is seen 
approaching a layering transition. 
Several protocols have been followed together with different cells. 
We have shown that it exists a true hysteresis loop which persists 
when the sampling time of the experiment is increased 
and which depends on the geometry of the experiment. 
We have also seen that sustained oscillations can be observed 
but in a metastable way. A very rich signal can also been obtained 
which can eventually be seen as chaotic. In what follows we will 
address first the nature of the temporal dependences. 
More specifically we want to understand how much these temporal dependences are related to textural changes. 
Then, we will try to address the question of the nature of the complex signal (whether it is chaotic or not).  

\section{A coupling between temporal instabilities, structure  and spatial instabilities}

We have seen that complex dynamics can be described in this system. A previous work  
has shown that the observed oscillations of  
shear rate were correlated to structural changes
in this complex fluid system \cite{Wunenberger:01}~: the fluid oscillates between 
high shear rate values corresponding to the layered state and low shear 
rate values corresponding to the disordered state.
In fact, there are also macroscopic instabilities that occur in complex fluids : bands 
in the vorticity direction and in the  $\nabla\!v$ direction may appear in the vinicity 
of out-of-equilibrium transitions (vorticity and shear banding). 
These instabilities have been extensively studied 
both theoretically and experimentally
\cite{Spenley:93,Berret:97,Fisher:01,Goveas:01}. These macroscopic 
instabilities may lead to a coupling with the dynamical observed behavior in our system.
In this Section, we report arguments to show that the vorticity direction is irrelevant 
for the dynamical recorded shear rate and we suggest that coupling with spatial structures 
in the $\nabla\!v$ direction may occur.

\begin{figure}[ht]
\begin{center}
\epsfxsize= 8.0cm
\centerline{\epsfbox{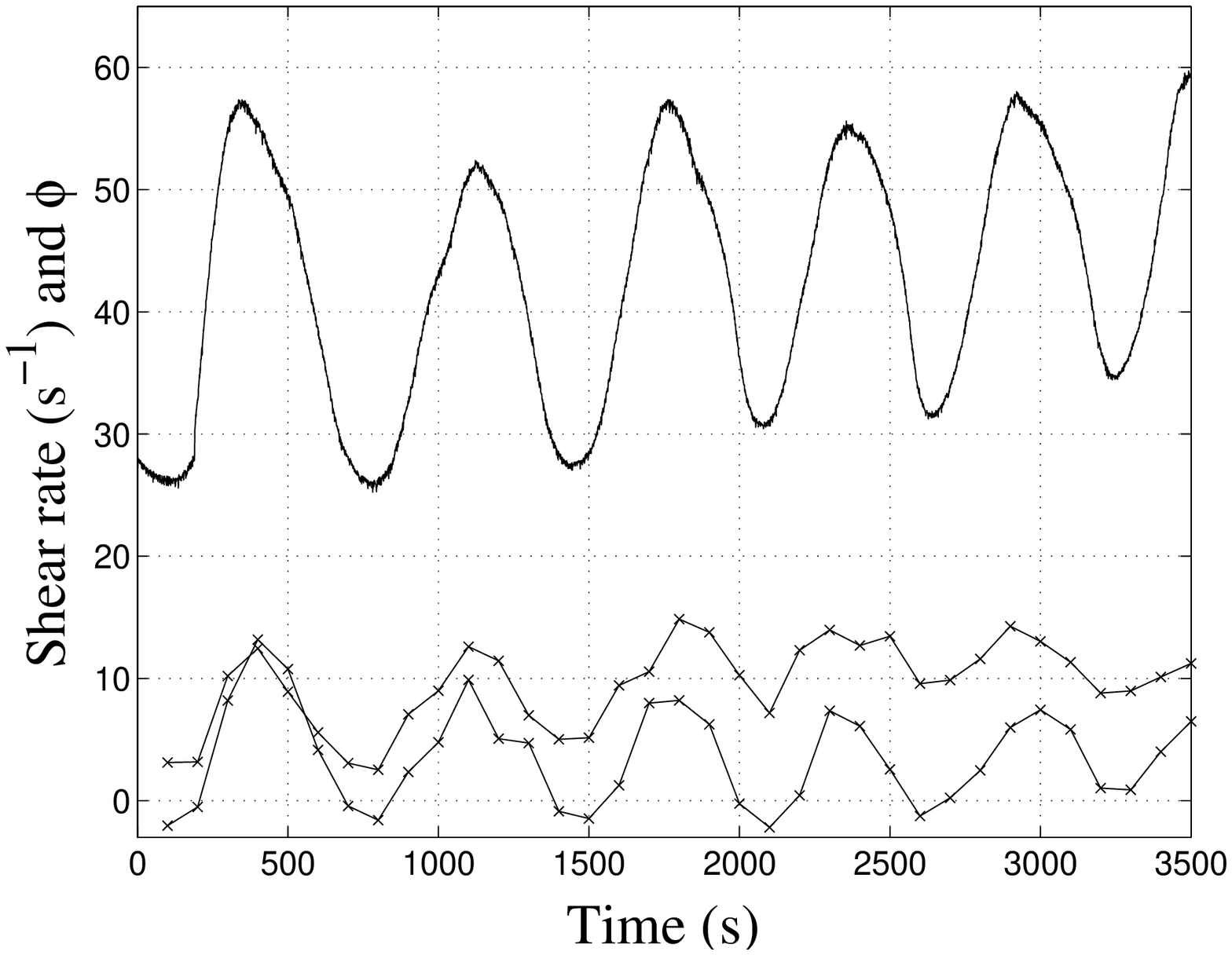}}
\end{center}
\caption{Correlation between two contrast parameters $\phi$  at two different
heights in the Couette cell ($\times$) with a gap $e = 0.5$~mm
and $\dot{\gamma}(t)$ (solid line) 
in region 4.}
\label{correlation2height}
\end{figure}

Observing the sample with naked eyes allows us to see quite a lot of inhomogeneities
in the vorticity direction. These inhomogeneities can be described as horizontal
bands. 
These bands have a weak contrast and delimit the Couette 
cell into different regions of turbidity.
There is also no selection of a wavelength,
and thus no systematic
number or size of bands have been seen. The typical range of observed size is 0.1~mm to 10~mm.
These bands appear systematically during transient phases and rarely in asymptotic states.
They appear roughly at the level of point B and persist
until point F. The dynamics of these bands seems not correlated to the dynamical recorded viscosity. 
To understand better the coupling between these macroscopic structures and 
rheology, 
we decide to use the diffraction patterns recorded on the CCD (cf. Fig.~\ref{setup}) 
at two different heights in the cell.
The direct beam of the laser is hidden by a beam stop to avoid the saturation
of the signal.
A contrast
parameter $\phi$ can be defined. It is convenient to define it as naught 
in the high symmetry texture 
(disordered state of onion) and different from zero in the low symmetry texture
(layered state). We chose to define $\phi$ as the
difference between the mean intensity of scattered
light in a region of the ring where a peak appears and the mean
intensity on a region of the ring where no peak ever appear (cf. Fig.~\ref{diffraction}).
In a region where no structural
changes occur, we checked that $\phi$ does not depend on 
the shear rate $\dot{\gamma}$.
In fact below point B and above point E, the stationary states correspond 
respectively to a ring of scattering ($\phi = 0$) and a modulation of scattered light on the ring 
($\phi > 0$).  
Figure \ref{correlation2height} shows contrast parameters $\phi$,
measured at two different heights $z_1$ and $z_2$ ($\|z_1 - z_2\| \approx 1$~cm)
in the Couette cell and the corresponding oscillating shear rate in
region 4. The similarity between the three time series is obvious.
The \textit{local}
measures of $\phi$ (only integrated into the $\nabla\!v$ direction) at two different heights,
are directly correlated to the \textit{global} measure of the shear rate. This
demonstrates that the shear rate oscillations, if coupled with
macroscopic spatial structures, are invariant under a $z$-translation in the Couette cell 
even though inhomogeneities can be observed.
The results shown in Fig.~\ref{correlation2height} are (i)
the existence of correlations between the oscillating viscosity and the microstructure of the phase,
(ii) their presence in all the height of the cell. 
Indeed we tried different Couette cells presenting different heights (30 and 10~mm).
No spectacular changes were observed in the temporal dependence of the signal and 
the observed dynamical scenario 
is still present.  
These observations are still comforting us, in the fact that, vorticity
is (at first approximation) irrelevant for the dynamical scenario we observe. 
The change in the gap was by far more spectacular.

\begin{figure}[ht]
\begin{center}
\epsfxsize= 8.0cm
\centerline{\epsfbox{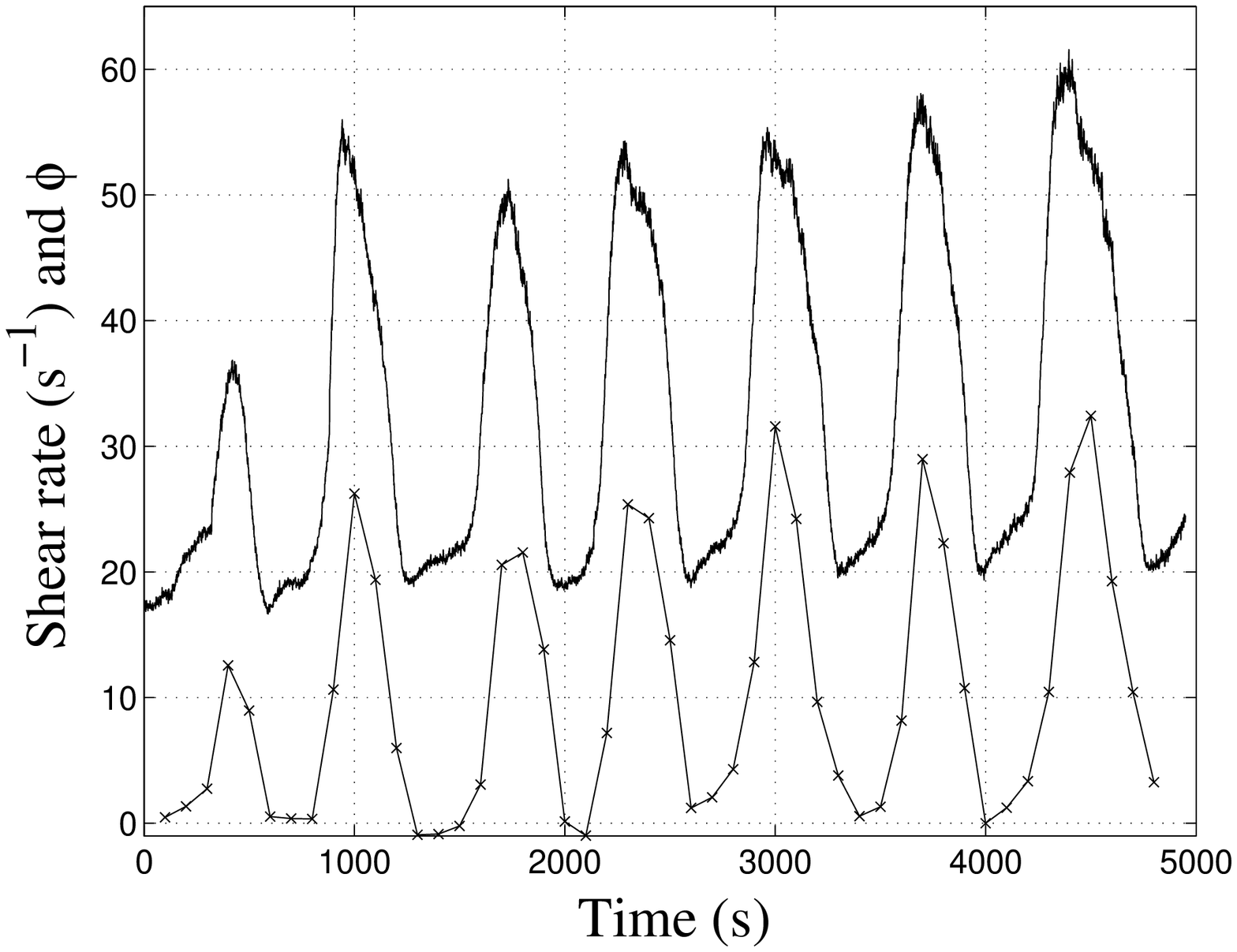}}
\end{center}
\caption{Correlation between the contrast parameter $\phi$ ($\times$)
 and $\dot{\gamma}(t)$ (solid line) 
in region 4 with gap $e=0.5$~mm.}
\label{correlation}
\end{figure}

\begin{figure}[ht]
\begin{center}
\epsfxsize= 8.0cm
\centerline{\epsfbox{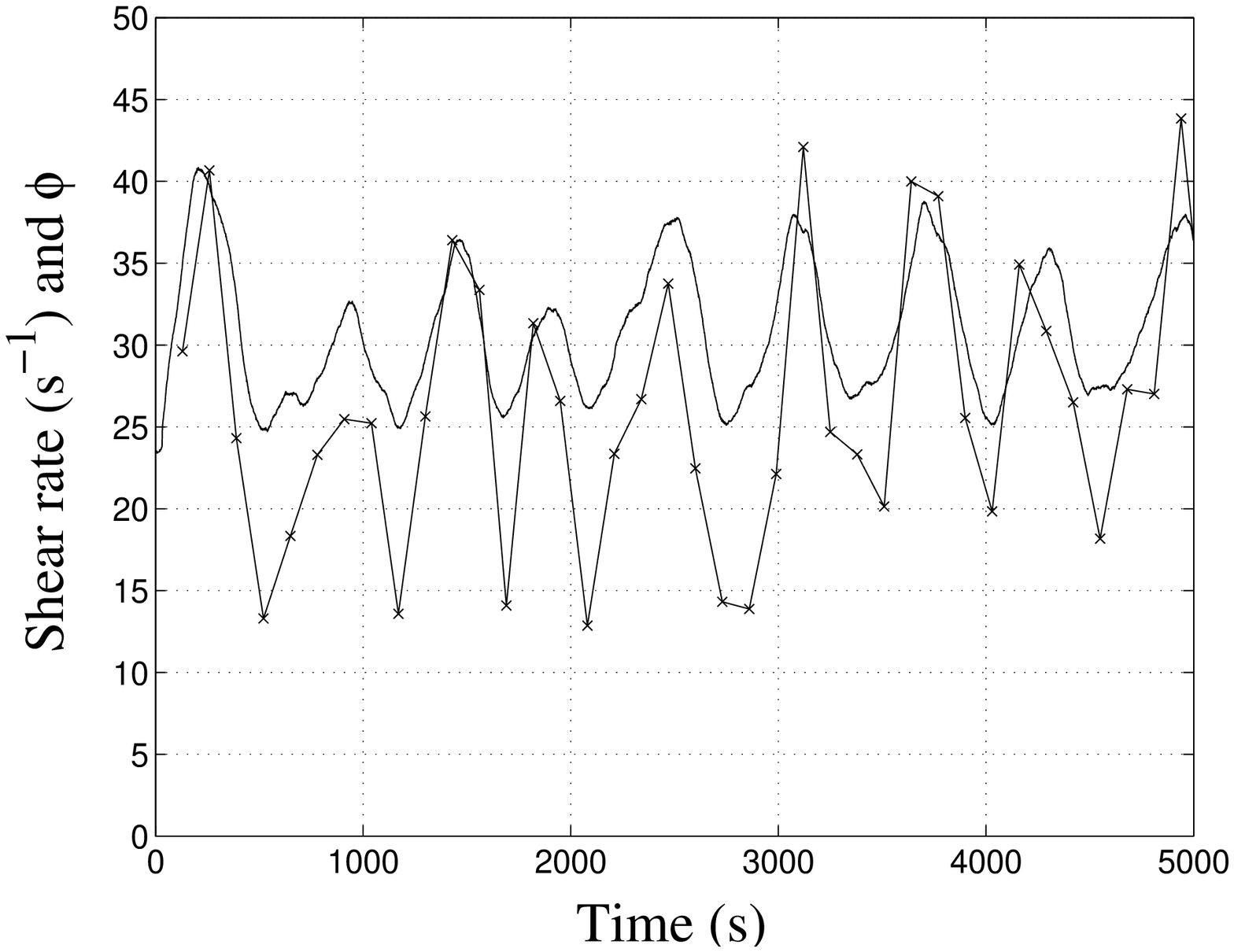}}
\end{center}
\caption{Correlation between the contrast parameter $\phi$ ($\times$) 
and $\dot{\gamma}(t)$ (solid line) 
in region 4 with gap $e=1$~mm.}
\label{correlation2}
\end{figure}

In the precedent Section, rheological measurements were 
performed in two different geometries with
different gaps.
A confinement in the gap may change the observed dynamics 
if spatial structures lying in the
$\nabla\!v$ direction were oscillating. In fact the flow curves with $e=0.5$~mm are quite similar 
to the flow curves with $e=1$~mm and the mean periods of the recorded
oscillations are equivalent, but some rheological differences were noticed above. 
There is a new difference when correlation with microstructure is done. 
The large oscillations observed in a non-asymptotic 
protocol with $e=0.5$~mm are directly correlated with the contrast parameter, 
as shown in Fig.~\ref{correlation}. 
The complex fluid oscillates between the two branches 
(disordered state and layered state)
since $\phi$ oscillates between zero (isotropy of the ring) and a non-zero value 
(peaks on the ring)~: region 4 corresponds to homogeneous relaxational oscillations for the small gap. 
For the larger geometry ($e=1$~mm), there is also correlation between structure 
and flow as shown in
Fig.~\ref{correlation2}. However, the contrast parameter does not oscillate sharply between zero
and a non zero value~: there is already an anisotropy of the ring in this region. The system 
oscillates between a state where spatial correlations are weak (low shear rate values) to a state
where the latter are strong (high shear rate values).
In the larger gap, the region 4 corresponds to 
oscillations between a mixture of these states. So this may correspond to a separation 
in different oscillating ordered structures in the $\nabla\!v$ direction.
If such spatial structures in the $\nabla\!v$ direction were existing, the flow curves will
depend on the gap since spatial effects between oscillating structures may be stronger in the large gap 
than in the small gap. 
This may  explain the observed difference between $F \rightarrow G$ and 
$G \rightarrow F$ in the flow curve in Fig.~\ref{flowcurve1gap}.
The schematic flow curve presented in Fig.~\ref{scenario} correspond 
to a homogenous case, well fitted
by the real flow curve measured with $e=0.5$~mm. 
These observations suggest that the 
$\nabla\!v$ direction is relevant for the dynamical observed scenario.

To summarize the results of this section,
there are arguments to show that the observed dynamical complexity does not depend
(at first approximation) on the vorticity direction. It 
depends more on the $\nabla\!v$ direction. However the strong differences between
the two geometries
($e=1$~mm and $e=0.5$~mm) are observed only in the metastable region.
In particular, to the precision of the device ($\delta t \approx 15000$~s and
$\delta\sigma \approx 0.05$~Pa), no
simplifications of the dynamics of the viscosity were observed  with the small gap.
We will come back on 
these results in the Sec.~IV.

\section{Analysis of the dynamical behavior of viscosity with dynamical system theory}

At this stage, we would like to spend some time to analyse the observed experimental
behavior, following the out-of-equilibrium theories which have been
experimentally and theoretically developed in the last 30~years.
In particular
the complex signal that we observed on branches ($B \rightarrow D$ and $E \rightarrow B$)
could eventually be described with low dimensional dynamical systems 
as recently suggested 
on other complex fluids \cite{Bandyopadhyay:00}.
To assess low dimensional chaotic properties, there exist mainly three techniques based 
on the properties of low dimensional dynamical systems~: 
(i) to exhibit the \textit{transition to chaos},
(ii) to exhibit the deterministic application which creates a strange attractor, 
(iii) to compute the metric invariants of a strange attractor \cite{Gilmore:98}. 
The latter method has been used 
in the following works \cite{Bandyopadhyay:00}
 in the context of rheology of wormlike micelles, 
but this method required very long time series 
to give good estimates of the metric invariants 
(such time cannot be reached in typical rheological experiments),
moreover there are no theories
which can give errors for these invariants \cite{Gilmore:98}, and finally 
this method do not provide the topological properties
of a strange attractor. 
For these reasons, we have followed the two 
other methods to analyse our data.       
For that we need to recall some basic principles of dynamical system theory and
we will use these principles as a framework to analyse our data. 

Dynamical system theory describes the properties of solutions of dynamical systems which are
sets of Ordinary Differential Equations (ODE). A dynamical system can be written as
\begin{equation}
\dot{X} = f(X,\mu)
\end{equation}
where $X$ is a n-dimensional vector which evolves according to (1). $\mu$ is a
p-dimensional vector of parameters which controls the mathematical form of the function $f$.
The solutions of Eq.~(1) can be seen as trajectories in a
n-dimensional space called the 
\textit{phase space}. These trajectories 
cannot cross each other due to the unicity of solution for a 
given initial condition. In a lot of physical systems, dissipation of energy occurs, 
this characterizes a \textit{dissipative} system. In such systems, 
after a transient phase, all the trajectories collapse on a subspace $\mathcal A$, called the 
\textit{attractor}. The dimension of this attractor has the following property due to
dissipation~: $d(\mathcal{A}) <  n$.
The topological properties of the attractor  are of 
major importance to study the \textit{asymptotic} 
solutions of a dissipative dynamical system.  

Dissipative dynamical systems of dimension 3 may  
exhibit solutions which are aperiodic \cite{Lorenz:63}.
Such solutions are extremely sensitive
to initial conditions and their dynamics cannot be predicted. Such solutions 
are called \textit{chaotic} solutions. In low dimensional dissipative 
dynamical systems with $n = 3$,
which exhibit chaotic solutions, the mathematical condition $2 < d(\mathcal A) < 3$ 
for the dimension of the attractor is required. So such an attractor has \textit{fractal} 
properties and is called \textit{strange attractor}. 

In the theory of dynamical systems, the solution evolves from a stationary state (i.e. $\dot{X} = 0$)
to a chaotic state following a set of bifurcations as $\mu$ is changed. A bifurcation is the 
passage from a solution to an other which is not topologically equivalent to the first 
\cite{Guckenheimer,Crawford:91}.
The set of bifurcations necessary to create a strange attractor is called the 
\textit{transition}
to chaos. Experimentally, to show such a transition by changing the parameters 
$\mu$ of the experiment
is a strong proof for the existence of a chaotic state. 

In a lot of physical systems, the presumed model which reproduces the dynamics may 
have a high number of equations. 
However when approaching a bifurcation, 
the \textit{normal form theorem} may allow to reduce the complexity of the 
equations to a simple equation called the \textit{normal form} which described
the dynamics in the vinicity of the bifurcation. Since the bifurcations are not 
generically simultaneous, the dynamics can be reduced to a low dimensional dynamical system
with dimension increasing from 1 as the control parameters are changed. 
This theorem
involves that for complex systems (like rheology of complex fluid), we may expect that the 
dynamical aperiodic states near an out-of-equilibrium transition,  
can be expressed as the solutions of a 3-dimensional dynamical system.

\begin{figure}[ht]
\begin{center}
\epsfxsize= 8.0cm
\centerline{\epsfbox{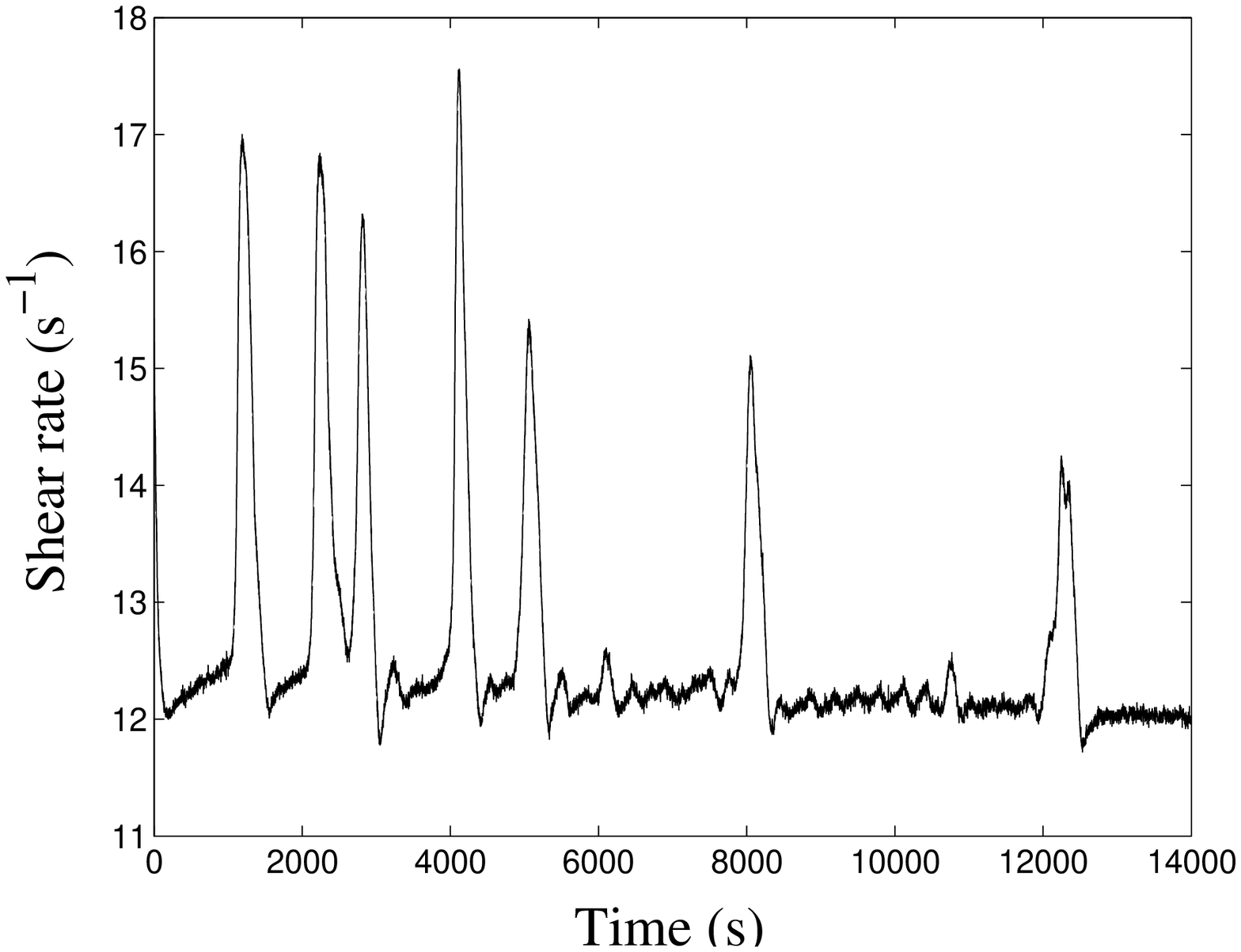}}
\end{center}
\caption{Typical time serie showing the simplification 
of the dynamics of region 6 when approaching $\sigma_0$.}
\label{simplification}
\end{figure}

\begin{figure}[ht]
\begin{center}
\epsfxsize= 8.0cm
\centerline{\epsfbox{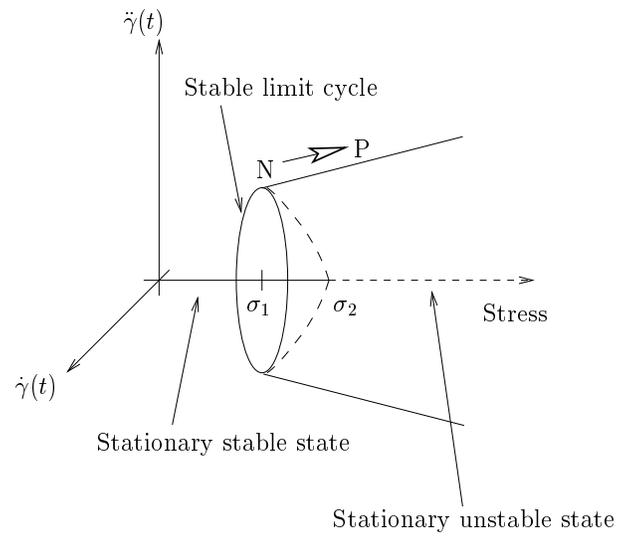}}
\end{center}
\caption{Subcritical Hopf bifurcation and transition to chaos, $\sigma_2$ corresponds 
to the lost of stability of the stationary
state. At stress $\sigma_1$, a limit cycle exists, but between N and P,
transition to chaos may occur}  
\label{hopf}
\end{figure}

In the experiment described above, two parameters $\mu$ can be used : the stress $\sigma$
and the temperature $T$.  
Previously different regims of dynamical behavior of the viscosity have been presented.
Namely the aperiodic oscillations recorded in the vinicity of the point C 
(Fig.~\ref{aperiodic}) may be described with a 3-dimensional dynamical chaotic system. 
As discussed above, if this dynamics corresponds to a 
chaotic state, a transition to chaos should be present, in particular a Hopf bifurcation leading
to a limit stable cycle should exist (i.e. a periodic state). 
This Hopf bifurcation may lead to the behavior of the 
Fig.~\ref{sixregion}(b), where before any transition, the stationary state becomes noisy 
with a period which is the 
same as in the aperiodic state. This phenomenom is called \textit{stochastic resonance} and 
corresponds to the amplification of noise near a Hopf bifurcation \cite{Gang:93}.
In fact, at the precision of the device ($\delta\sigma \approx 0.01$ Pa,
$\delta t \approx 15000$~s and
$\delta T = 0.1^\circ$C) when stress or temperature is varied no periodic asymptotic viscosity 
has been recorded before this aperiodic state. So if a Hopf bifurcation exists, 
which is necessary 
to create a chaotic dynamics, it must be subcritical. Such a case is shown on Fig.~\ref{hopf}, 
where the chosen parameter is stress. When $\sigma < \sigma_1$ the stationary state is stable, 
there is just stochastic resonance when stress approaches $\sigma_1$ (region 2).
 At stress $\sigma_1$, 
a stable limit cycle appears with a finite amplitude.
Between $\sigma_1$ and $\sigma_2$ the stable stationary state and the 
stable limit cycle coexist, but some 
bifurcations may arise on the cycle which lead to a chaotic dynamics ($N \rightarrow P$). When 
$\sigma > \sigma_2$ the stationary state is no longer stable and the stable asymptotic state 
corresponds to a chaotic state. 
This scenario could eventually correspond to the experimental observed one~: region 1 
$\rightarrow$ region 2 $\rightarrow$ region 3. In order to assess this hypothesis, 
when stress oscillates 
aperiodically at $\sigma \geq \sigma_2$ in region 3, when stress goes down on the branch 
P $\rightarrow$ N, we may expect simplification to a periodic state. To the precision 
$\delta\sigma \approx 0.01$~ Pa and $\delta t \approx 15000$~s, no periodic viscosity has been recorded.
This observation means 
that $\|\sigma_1 - \sigma_2\| \leq 0.01$~Pa.
Such an observation has been also checked with temperature as parameter.
As discussed above in Sec.~III, no simplier dynamics arise when approaching point B when 
stress goes down from region 6. The only transition we observed
is the recorded time serie shown in
Fig.~\ref{simplification}. So, for the transition $G\rightarrow B\rightarrow A$, 
no Hopf bifurcation
has been seen. 
However the dynamical responses of the shear rate 
in the vinicity of point B (Fig.~\ref{simplification}) 
and C (Figs.~\ref{longtime} and \ref{sixregion}(b)) strongly 
suggest the presence of  subcritical Hopf bifurcations.

Since no success in showing the scenario related to a Hopf bifurcation has been obtained,
one could eventually try to analyse the signal obtained in the aperiodic regime.
For that we can use techniques which have been developed to demonstrate 
the chaotic nature of experimental data. 
There are a lot of methods to assess whether a recorded time serie is chaotic of low dimensionality.
They involve
different invariants~: metric, dynamic and topologic \cite{Gilmore:98}. 
The two first methods compute 
the metric and dynamic invariants of a strange attractor, such as Lyapunov exponents and
various dimensions of the strange attractor. 
No statistical theory exists that assigns errors to the latter, so it is impossible to
determine the validity of the computed invariants. 
In our case, the time series recorded in region 3 contain up to 80 oscillations, so it is 
impossible to use those methods. 
The third method is based on the topological properties of the strange attractor. Currently
100 oscillations are enough to assess chaotic dynamics. Moreover the sampling
time intervall is 1~s in the recorded time series, 
which leads to 500 points per cycle~:  this is 
enough to use the method. Strange attractors are topological 
objects with fractal properties, 
which allow to have the \textit{sensitive to initial conditions} property 
between two trajectories. To compute a strange attractor using a time serie, 
one may use the embedding theorem. Such a theorem conjectures that, 
for a dynamical system like Eq.~(1)
the attractor $\mathcal A$ constructed with the natural variables $\{X_i(t)\}$ 
is topologically equivalent to the construction of $\mathcal A$ with the following variables~: 
$\{X_i(t), X_i(t+\tau), X_i(t+2\tau),\ldots\}$. This is called the \textit{time delay embedding}. 
The delay time $\tau$ is arbitrary but a useful choice must
be found to study an experimental time serie. Other embedding variables can be used like~: 
$\{X_i(t), \dot{X_i}(t), \ddot{X_i}(t),\ldots\}$ and other combinaisons.
The embedding theorem allows when just one variable is measured as in a lot of 
experimental devices, to reconstruct the 
attractor $\mathcal A$ without knowing all the variables of the dynamical system. 

\begin{figure}[ht]
\begin{center}
\epsfxsize= 8.0cm
\centerline{\epsfbox{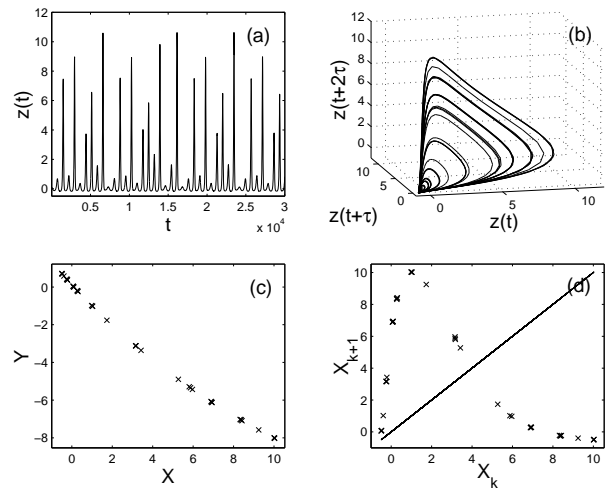}}
\caption{(a) Solution z(t) of the R\"{o}ssler system, (b) Embedding reconstruction of the attractor,
(c) Poincar\'{e} section (d) First return map constructed with 
the abscissa X of the Poincar\'{e} section}
\label{rossler}
\end{center}
\end{figure}

Let us first present a typical case, we will use this case in a modified way 
at the end of the article.
Such a construction is
given for the R\"{o}ssler system, with $(a,b,c) = (0.3,0.3,4.5)$~: 
\begin{equation}
\begin{cases}
\dot{x} &= -y-z  \\
\dot{y} &= x+ay  \\
\dot{z} &= b + z(x-c)
\end{cases}
\end{equation}
Such a system exhibits chaotic solutions for the given parameters. 
The numerical integration 
of this system is shown in Fig.~\ref{rossler}(a) where the variable $z(t)$ is reproduced.
In Fig.~\ref{rossler}(b) is shown the embedded corresponding 
attractor with a delay time $\tau= 20$, which 
corresponds to $\tau \approx \frac{T_0}{40}$, where $T_0$ is the mean period of the signal. To 
show deterministic chaos, one may find the deterministic application which generates 
the strange attractor. This topological approach is based on the study of the Poincar\'{e} section
which corresponds to the intersection of the trajectories lying on the attractor $\mathcal A$ 
and a plan. Such a Poincar\'{e} section is constructed and plotted 
in Fig.~\ref{rossler}(c). The chosen Poincar\'{e} section corresponds to the plan defined 
by the normal vector $(-1,0,1)$ in the frame $[z(t), z(t+\tau), z(t+2\tau)]$. 
The Poincar\'{e} section is a line and this corresponds to the dissipation 
of the R\"{o}ssler system. When $X_{k+1}$ vs $X_{k}$ are plotted where $X_k$ corresponds to the 
abscissa on the Poincar\'{e} section of the k-intersection of the trajectory with the latter, (shown 
in Fig.~\ref{rossler}(d)), the corresponding curve has a determined shape with a single maximum and 
a slope which is greater than 1 at the intersection with the bissecting line. This is called 
the \textit{first return map}. The time serie shown in Fig.~\ref{rossler}(a) does not seem to be 
predicted, 
but the construction of the attractor $\mathcal A$, the Poincar\'{e} section and finally the  
first return map reveal the deterministic application which is characteristic of deterministic
dissipative chaotic dynamical system. The shape of the first return map
allows to predict the (k+1)-intersection of the trajectory with 
the Poincar\'{e} section if the k-intersection
is known : this is the deterministic property of the equations. 
However the single humped shape of the first return map,
with an average slope greater than one, 
permits to have the sensitive to initial conditions property between two trajectories.   
 
\begin{figure}[ht]
\begin{center}
\epsfxsize= 8.0cm
\centerline{\epsfbox{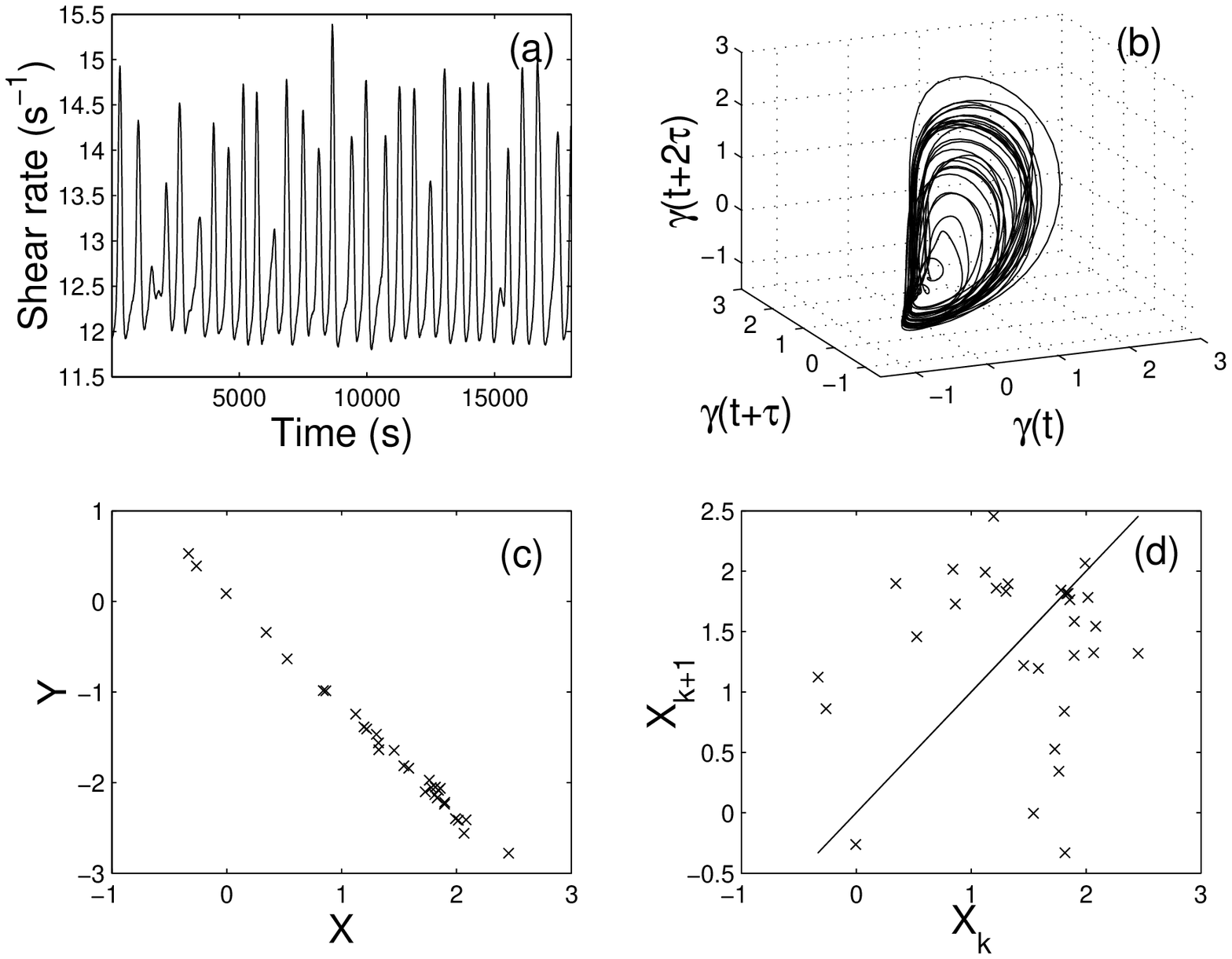}}
\caption{(a) Analysis of the experimental time serie, (b) Embedding 
reconstruction of the attractor, (c) Poincar\'{e} section, 
(d) First return map constructed with the abscissa X of the Poincar\'{e} section}
\label{analysis}
\end{center}
\end{figure}

In the experimental time series recorded, we have followed carefully
the same analysis. Premilinary, a high 
frequency filter has been used to eliminate the noise due 
to the frequency of the rotation of the Couette cell.
This noise has an amplitude of about $0.05$~s$^{-1}$ and 
corresponds to the high frequency of the rotation of the 
geometry $(10 \ll 500$~s). This procedure leads to the time serie presented 
in Fig.~\ref{analysis}(a). The embedded attractor is constructed using the time delay method, with
a delay time $\tau = 40$ s (Fig.\ref{analysis}(b)).
The constructed attractor is qualitatively similar to the R\"{o}ssler's one.
So the recorded time serie in region 3 is similar to a chaotic variable 
of a 3-dimensional dynamical system. To assess this property   
Poincar\'{e} section is plotted in
Fig.~\ref{analysis}(c) and corresponds to the plan with normal vector $(-1,0,1)$. However
the first return map shown in Fig.~\ref{analysis}(d) exhibits no simple shape. Other Poincar\'{e} 
sections and other choices of $\tau$ have been investigated, but no simplier shape 
has been found. We also varied the filter and defined curvilinear abscissa on the Poincar\'{e} section
in order to reconstruct the first return map, but no deterministic application has been found.

As a conclusion of this work, we cannot assess that the recorded time series of the shear rate 
correspond to dissipative deterministic chaos of dimensionality 3, even though a great
similarity. This might be, 
because the stastistic
we study is very poor~: up to 40 oscillations have been studied. This might due also, 
to a low noisy frequency dynamics. The studied dynamics may also correspond to a 4-dimensional chaotic
state but in this case the transition to the latter requires the presence of 
a 3-dimensional strange attractor which has not been observed. 
However we cannot exclude that this result suggests 
that the observed dynamics may be more complex. A few coupling effects 
with space may occur and so a spatio-temporal dynamics may be recorded as discussed in Sec.~III.
Such a case does not lead to a simple shape in the first 
close return map as we will see later.
Actually rheology is an experimental tool used for probing the viscosity 
of materials. However the measure is only global~: 
the recorded shear rate does not correspond to the local shear rate in the gap. 
The same occurs for the control parameter~: a torque is imposed on the axis of the rheometer, 
but the local stress induced in the fluid is not known. This two phenomena lead to spatial 
structures like shear banding and vorticity banding in complex fluid as discussed in Sec.~III.

\begin{figure}[ht]
\begin{center}
\epsfxsize= 8.0cm
\centerline{\epsfbox{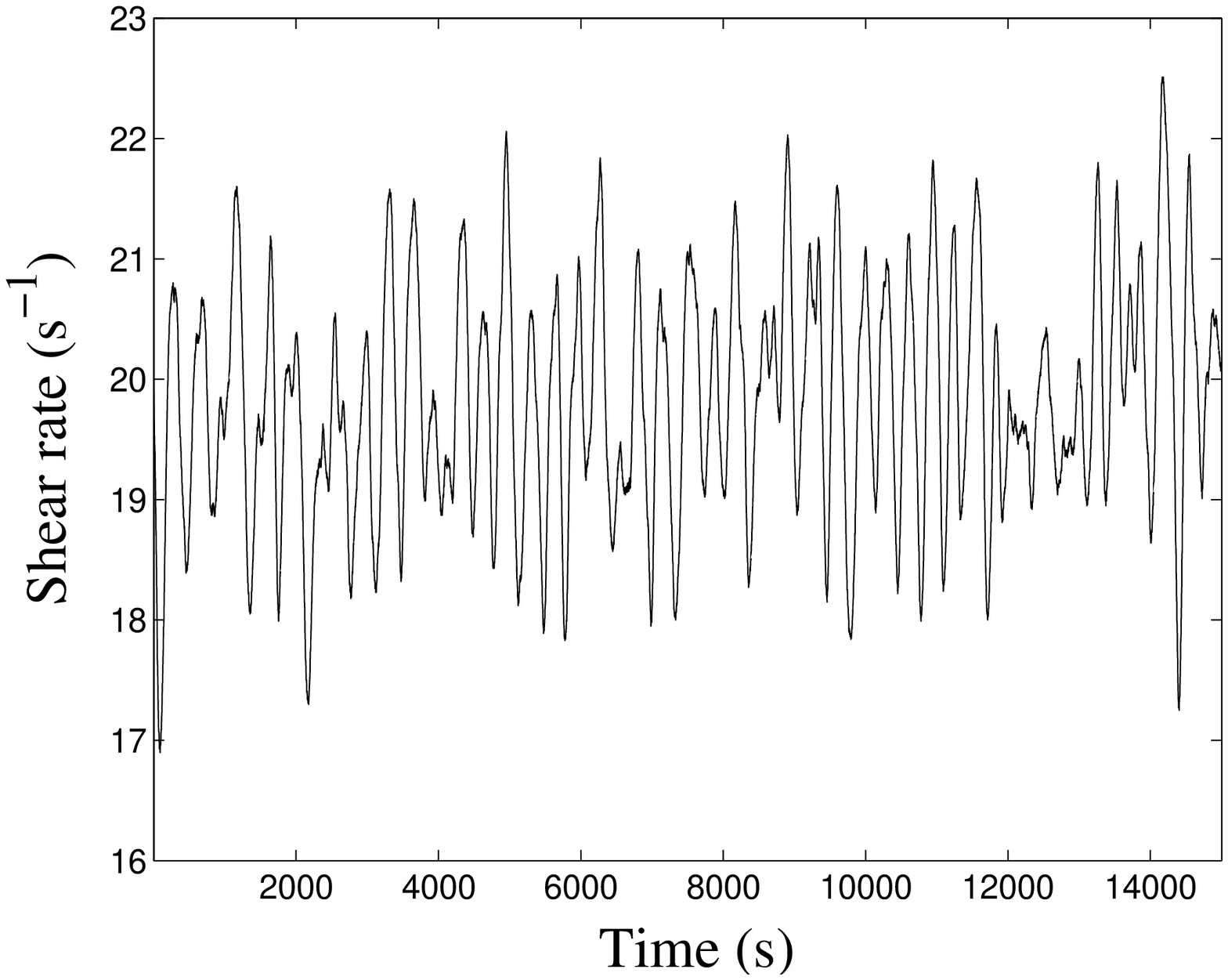}}
\end{center}
\caption{Complex dynamical time serie $\dot{\gamma}(t)$ in region 6.}
\label{region6}
\end{figure}

\begin{figure}[ht]
\begin{center}
\epsfxsize= 8.0cm
\centerline{\epsfbox{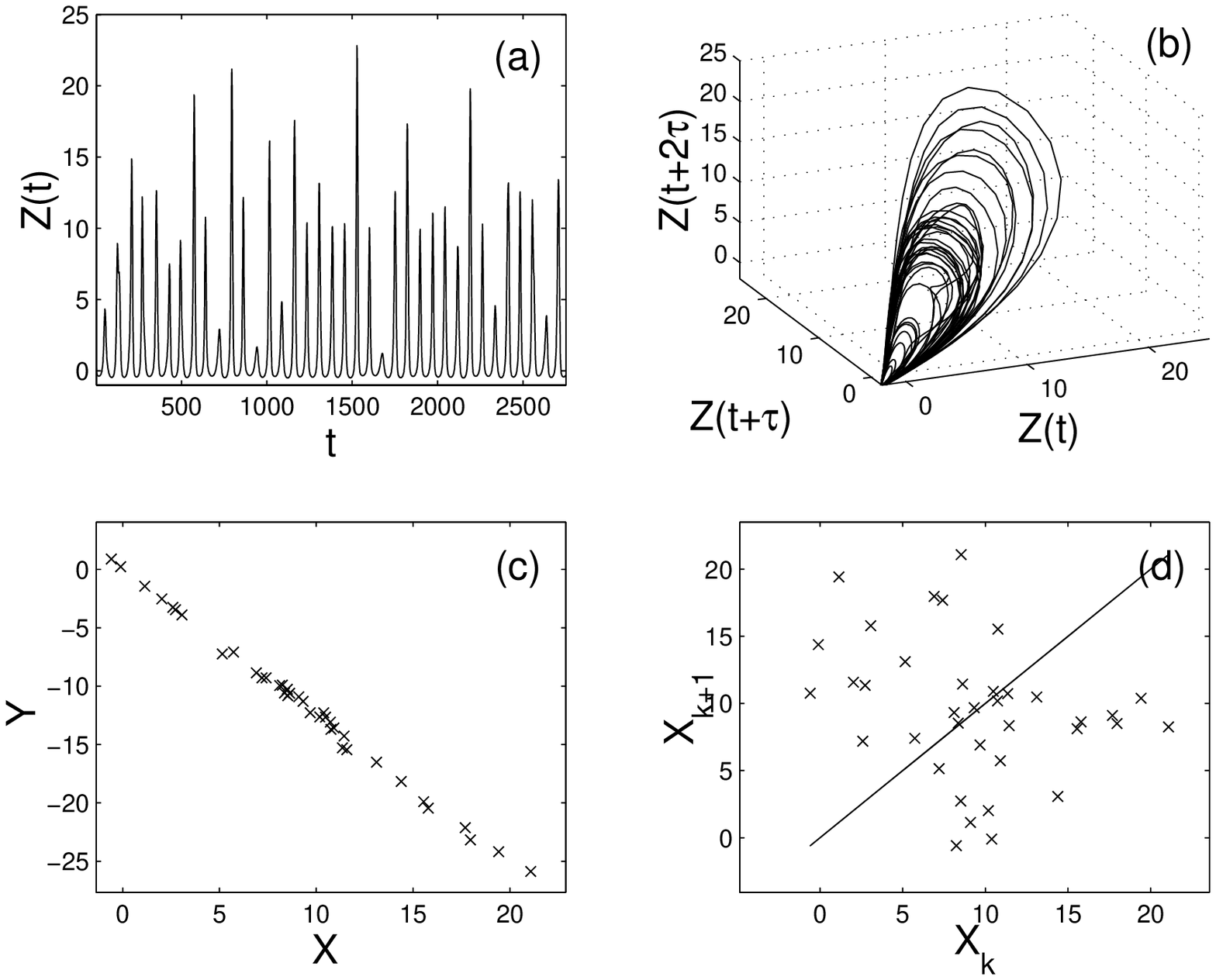}}
\caption{(a) Analysis of the time serie $Z(t)$, (b) Embedding 
reconstruction of the attractor, (c) Poincar\'{e} section, 
(d) First return map constructed with the abscissa X of the Poincar\'{e} section}
\label{threerossler}
\end{center}
\end{figure}

In order to illustrate 
this kind of effects on a 3-dimensional dynamical system, we present here
a simple illustration with the help of the R\"{o}ssler system. In a case where local 
spatial structures are oscillating, the rheological global measure corresponds to the sum
of these local structures.    
Three integrations 
of the R\"{o}ssler system with $(a,b)=(0.3,0.3)$ and $c = 4.5; 4.4$ and $4.7$ have been made. 
These variables 
$\{z_1(t), z_2(t), z_3(t)\}$ may correspond to local 
chaotic oscillating structures.
The sum $Z(t)$ of the 
three different variables $z_1(t)$, $z_2(t)$ and $z_3(t)$,
plotted in Fig.~\ref{threerossler}(a), may correspond to the global 
measure of rheological experiments. 
The topological reconstruction of the first return map 
is the same as presented above and shown in Fig.~\ref{threerossler}(b), (c) and (d).
The construted attractor has a similar shape as the R\"{o}ssler's attractor, however
no deterministic
shape is found for the first return plot. So the dynamical variable 
$Z(t)$ does not simply corresponds to 
dissipative deterministic chaos of dimensionality 3. This illustrates that a few spatial effects 
could lead to the results presented concerning the recorded shear rate. 
Actually this example is very simple, spatial coupling effects may occur between these 
three variables, but a more complicated case involving three oscillating variables with 
non linear coupling terms would provide the same result. 
In fact the illustration 
presented here does not prove these structures, it only suggests that a few spatial structures
may lead to aperiodic time series which are not strictly chaotic of 3-dimensionality
even though the recorded shear rate is qualitatively similar to a  3-dimensional chaotic
variable. 
In fact if more than three local variables where used to reconstruct a global measure, 
complex dynamical states as presented in Fig.~\ref{region6} can be reproduced easily. 
This region does not correspond to simple chaotic dynamic, 
but this may due to a lot of spatio-temporal effects.     

\section{Conclusion}
In this article we have presented a detailed experimental 
study of the dynamical behavior of the rheology of a complex fluid near a textural instability. 
We have shown that a complex behavior with different regims as a function of time can be described. 
Among the most interesting regimes, sustained oscillations and \textit{chaotic like} types of 
signals have been observed. We have shown that the temporal dependence is related to textural changes 
involving the whole sample. We spend sometime to analyse the \textit{chaotic like} signal, 
using a careful mathematical analysis. We cannot prove that the signal corresponds 
really to a 3-dimensional chaotic system, even though it has several distinctive features 
resembling to a 3-dimensional deterministic chaotic state.
To conclude, to interpret the signal 
we have, we make the hypothesis that there is a coupling between temporal behavior and 
spatial instabilities involving a finite but small number of cells. We have also 
shown with rheophysics tools that such spatial structures are probably 
in the $\nabla\!v$ direction. 
A question remains about
the microscopic origin of the observed dynamics, it is obviously a complicated problem. However 
we can make some assumptions according to the experimental results discussed in this Article. First
the observed period of oscillations involves long time scales (about 10 minutes), such time 
scales have been previously observed with the same system in other experiments. Namely 
it has been shown that the structural response of the onion texture involves time scales of the order of minutes. 
This behavior was assumed to be related to the displacement of the grain boundaries of the 
disordered onion texture \cite{Panizza:96}. In an other context it has been measured with neutron scattering
that the smectic period of the ordered onion texture decreases with increasing shear rate \cite{Diat:95}.
It was assumed that the expelled water was staying between the different layers of the 
ordered state. When the shear rate is stopped, the swelling kinetics of the compressed onion texture shows 
strong nonlinear effects on long time scales \cite{Leng:01}. 
Finally, the time scales of reorganization of the onion texture
between two different imposed stress are of the order of minutes \cite{Panizza:98}.   
The oscillations may be the result of a competition between an ordering of the disordered state driven by the stress 
(mechanical ordering) and an slow textural evolution which destroys the stress-induced ordered state.    
These two effects may take place on different time scales and may produce oscillating behavior.
Concerning the strong dependence of the observed dynamics with temperature, it may be explained by
the dependence on temperature of the time scales discussed above.
For example it has been shown that the time scale of the swelling kinetics depend 
strongly on the temperature, due to the presence of thermally activated defects in the lamellar phase \cite{Leng:01}.

\acknowledgments{
The authors are deeply grateful to Alain Arn\'eodo for impulsing a lot of ideas in this work. 
We also thank S\'ebastien Manneville for fruitful discussions.
}

\pagebreak

\pagebreak
\newpage

\pagebreak
\newpage

\pagebreak
\newpage

\pagebreak
\newpage

\newpage

\begin{thebibliography}{45}
\expandafter\ifx\csname natexlab\endcsname\relax\def\natexlab#1{#1}\fi
\expandafter\ifx\csname bibnamefont\endcsname\relax
  \def\bibnamefont#1{#1}\fi
\expandafter\ifx\csname bibfnamefont\endcsname\relax
  \def\bibfnamefont#1{#1}\fi
\expandafter\ifx\csname citenamefont\endcsname\relax
  \def\citenamefont#1{#1}\fi
\expandafter\ifx\csname url\endcsname\relax
  \def\url#1{\texttt{#1}}\fi
\expandafter\ifx\csname urlprefix\endcsname\relax\def\urlprefix{URL }\fi
\providecommand{\bibinfo}[2]{#2}
\providecommand{\eprint}[2][]{\url{#2}}

\bibitem[{\citenamefont{Clark and Ackerson}(1980)}]{Clark:80}
\bibinfo{author}{\bibfnamefont{N.~A.} \bibnamefont{Clark}} \bibnamefont{and}
  \bibinfo{author}{\bibfnamefont{B.~J.} \bibnamefont{Ackerson}},
  \bibinfo{journal}{Phys. Rev. Lett.} \textbf{\bibinfo{volume}{44}},
  \bibinfo{pages}{1008} (\bibinfo{year}{1980}).

\bibitem[{\citenamefont{Safinya et~al.}(1990)\citenamefont{Safinya, Sirota,
  Plano, and Bruinsma}}]{Safinya:90}
\bibinfo{author}{\bibfnamefont{C.~R.} \bibnamefont{Safinya}},
  \bibinfo{author}{\bibfnamefont{E.~B.} \bibnamefont{Sirota}},
  \bibinfo{author}{\bibfnamefont{R.}~\bibnamefont{Plano}}, \bibnamefont{and}
  \bibinfo{author}{\bibfnamefont{R.~F.} \bibnamefont{Bruinsma}},
  \bibinfo{journal}{J. Phys. Condens. Matter} \textbf{\bibinfo{volume}{2}},
  \bibinfo{pages}{SA365} (\bibinfo{year}{1990}).

\bibitem[{\citenamefont{Diat et~al.}(1993)\citenamefont{Diat, Roux, and
  Nallet}}]{Diat:93}
\bibinfo{author}{\bibfnamefont{O.}~\bibnamefont{Diat}},
  \bibinfo{author}{\bibfnamefont{D.}~\bibnamefont{Roux}}, \bibnamefont{and}
  \bibinfo{author}{\bibfnamefont{F.}~\bibnamefont{Nallet}},
  \bibinfo{journal}{J. Phys. II France} \textbf{\bibinfo{volume}{3}},
  \bibinfo{pages}{1255} (\bibinfo{year}{1993}).

\bibitem[{\citenamefont{publishers}(1997)}]{Cambridge:97}
\bibinfo{editor}{\bibfnamefont{Applied Sciences Kluwer Academic} \bibnamefont{publishers}}, ed.,
  \emph{\bibinfo{title}{Theoretical challenges in the dynamics of complex
  fluids}} (\bibinfo{publisher}{Tom McLeish Series E}, \bibinfo{year}{1997}).

\bibitem[{\citenamefont{Cates and Evans}(2000)}]{Edimbourg:00}
\bibinfo{editor}{\bibfnamefont{M.~E.} \bibnamefont{Cates}} \bibnamefont{and}
  \bibinfo{editor}{\bibfnamefont{M.~R.} \bibnamefont{Evans}}, eds.,
  \emph{\bibinfo{title}{Soft and fragile matter~: Non equilibrium dynamics
  metastability and flow}} (\bibinfo{publisher}{Institute of Physics Publishing
  (Bristol)}, \bibinfo{year}{2000}).

\bibitem[{\citenamefont{Diat and Roux}(1995)}]{Diat:95_2}
\bibinfo{author}{\bibfnamefont{O.}~\bibnamefont{Diat}} \bibnamefont{and}
  \bibinfo{author}{\bibfnamefont{D.}~\bibnamefont{Roux}},
  \bibinfo{journal}{Langmuir} \textbf{\bibinfo{volume}{11}},
  \bibinfo{pages}{1392} (\bibinfo{year}{1995}).

\bibitem[{\citenamefont{Schmitt et~al.}(1994)\citenamefont{Schmitt, Lequeux,
  Pousse, and Roux}}]{Schmitt:94}
\bibinfo{author}{\bibfnamefont{V.}~\bibnamefont{Schmitt}},
  \bibinfo{author}{\bibfnamefont{F.}~\bibnamefont{Lequeux}},
  \bibinfo{author}{\bibfnamefont{A.}~\bibnamefont{Pousse}}, \bibnamefont{and}
  \bibinfo{author}{\bibfnamefont{D.}~\bibnamefont{Roux}},
  \bibinfo{journal}{Langmuir} \textbf{\bibinfo{volume}{10}},
  \bibinfo{pages}{955} (\bibinfo{year}{1994}).



\bibitem[{\citenamefont{Berret et~al.}(1997)\citenamefont{Berret, Porte, and
  Decruppe}}]{Berret:97}
\bibinfo{author}{\bibfnamefont{J.-F.} \bibnamefont{Berret}},
  \bibinfo{author}{\bibfnamefont{G.}~\bibnamefont{Porte}}, \bibnamefont{and}
  \bibinfo{author}{\bibfnamefont{J.-P.} \bibnamefont{Decruppe}},
  \bibinfo{journal}{Phys. Rev. E} \textbf{\bibinfo{volume}{55}},
  \bibinfo{pages}{1668} (\bibinfo{year}{1997}).

\bibitem[{\citenamefont{Pujolle-Robic and Noirez}(2001)}]{Pujolle:01}
\bibinfo{author}{\bibfnamefont{C.}~\bibnamefont{Pujolle-Robic}}
  \bibnamefont{and} \bibinfo{author}{\bibfnamefont{L.}~\bibnamefont{Noirez}},
  \bibinfo{journal}{Nature} \textbf{\bibinfo{volume}{409}},
  \bibinfo{pages}{167} (\bibinfo{year}{2001}).

\bibitem[{\citenamefont{Panizza et~al.}(1995)\citenamefont{Panizza,
  Archambault, and Roux}}]{Panizza:95}
\bibinfo{author}{\bibfnamefont{P.}~\bibnamefont{Panizza}},
  \bibinfo{author}{\bibfnamefont{P.}~\bibnamefont{Archambault}},
  \bibnamefont{and} \bibinfo{author}{\bibfnamefont{D.}~\bibnamefont{Roux}},
  \bibinfo{journal}{J. Phys. II} \textbf{\bibinfo{volume}{5}},
  \bibinfo{pages}{303} (\bibinfo{year}{1995}).

\bibitem[{\citenamefont{Panizza et~al.}(1996)\citenamefont{Panizza,
  Roux, Vuillaume, Lu and Cates}}]{Panizza:96}
\bibinfo{author}{\bibfnamefont{P.}~\bibnamefont{Panizza}},
  \bibinfo{author}{\bibfnamefont{D.}~\bibnamefont{Roux}},
\bibinfo{author}{\bibfnamefont{V.}~\bibnamefont{Vuillaume}},
\bibinfo{author}{\bibfnamefont{C.Y.D}~\bibnamefont{Lu}},
\bibinfo{author}{\bibfnamefont{M. E.}~\bibnamefont{Cates}}, 
  \bibinfo{journal}{Langmuir} \textbf{\bibinfo{volume}{12}},
  \bibinfo{pages}{248} (\bibinfo{year}{1996}).

\bibitem[{\citenamefont{Sollich et~al.}(1997)\citenamefont{Sollich, Lequeux,
  Hebraud, and Cates}}]{Sollich:97}
\bibinfo{author}{\bibfnamefont{P.}~\bibnamefont{Sollich}},
  \bibinfo{author}{\bibfnamefont{F.}~\bibnamefont{Lequeux}},
  \bibinfo{author}{\bibfnamefont{P.}~\bibnamefont{Hebraud}}, \bibnamefont{and}
  \bibinfo{author}{\bibfnamefont{M.~E.} \bibnamefont{Cates}},
  \bibinfo{journal}{Phys. Rev. E} \textbf{\bibinfo{volume}{78}},
  \bibinfo{pages}{2020} (\bibinfo{year}{1997}).

\bibitem[{\citenamefont{Onuki}(1987)}]{Onuki:87}
\bibinfo{author}{\bibfnamefont{A.}~\bibnamefont{Onuki}},
  \bibinfo{journal}{Phys. Rev. A} \textbf{\bibinfo{volume}{35}},
  \bibinfo{pages}{5149} (\bibinfo{year}{1987}).

\bibitem[{\citenamefont{Cates and Milner}(1989)}]{Cates:89}
\bibinfo{author}{\bibfnamefont{M.~E.} \bibnamefont{Cates}} \bibnamefont{and}
  \bibinfo{author}{\bibfnamefont{S.~T.} \bibnamefont{Milner}},
  \bibinfo{journal}{Phys. Rev. Lett.} \textbf{\bibinfo{volume}{62}},
  \bibinfo{pages}{1856} (\bibinfo{year}{1989}).

\bibitem[{\citenamefont{Ajdari}(1998)}]{Ajdari:98}
\bibinfo{author}{\bibfnamefont{A.}~\bibnamefont{Ajdari}},
  \bibinfo{journal}{Phys. Rev. E} \textbf{\bibinfo{volume}{58}},
  \bibinfo{pages}{6294} (\bibinfo{year}{1998}).

\bibitem[{\citenamefont{Olmsted and Goldbart}(1992)}]{Olmsted:92}
\bibinfo{author}{\bibfnamefont{P.~D.} \bibnamefont{Olmsted}} \bibnamefont{and}
  \bibinfo{author}{\bibfnamefont{P.~M.} \bibnamefont{Goldbart}},
  \bibinfo{journal}{Phys. Rev. A} \textbf{\bibinfo{volume}{41}},
  \bibinfo{pages}{4578} (\bibinfo{year}{1992}).

\bibitem[{\citenamefont{Bruinsma and Safinya}(1991)}]{Bruinsma:91}
\bibinfo{author}{\bibfnamefont{R.~F.} \bibnamefont{Bruinsma}} \bibnamefont{and}
  \bibinfo{author}{\bibfnamefont{C.~R.} \bibnamefont{Safinya}},
  \bibinfo{journal}{Phys. Rev. A} \textbf{\bibinfo{volume}{43}},
  \bibinfo{pages}{5377} (\bibinfo{year}{1991}).

\bibitem[{\citenamefont{Fischer}(2000)}]{Fisher:00}
\bibinfo{author}{\bibfnamefont{P.}~\bibnamefont{Fischer}},
  \bibinfo{journal}{Rheol. Acta} \textbf{\bibinfo{volume}{39}},
  \bibinfo{pages}{234} (\bibinfo{year}{2000}).

\bibitem[{\citenamefont{Wunenberger et~al.}(2001)\citenamefont{Wunenberger,
  Colin, Leng, Arn\'{e}odo, and Roux}}]{Wunenberger:01}
\bibinfo{author}{\bibfnamefont{A.-S.} \bibnamefont{Wunenberger}},
  \bibinfo{author}{\bibfnamefont{A.}~\bibnamefont{Colin}},
  \bibinfo{author}{\bibfnamefont{J.}~\bibnamefont{Leng}},
  \bibinfo{author}{\bibfnamefont{A.}~\bibnamefont{Arn\'{e}odo}},
  \bibnamefont{and} \bibinfo{author}{\bibfnamefont{D.}~\bibnamefont{Roux}},
  \bibinfo{journal}{Phys. Rev. Lett.} \textbf{\bibinfo{volume}{86}},
  \bibinfo{pages}{1374} (\bibinfo{year}{2001}).

\bibitem[{\citenamefont{Azkarate}(2000)}]{Cristobal:these}
\bibinfo{author}{\bibfnamefont{G.~C.} \bibnamefont{Azkarate}}, Ph.D. thesis,
  \bibinfo{school}{Universit\'{e} Bordeaux I} (\bibinfo{year}{2000}).

\bibitem[{\citenamefont{Bandyopadhyay et~al.}(2000)\citenamefont{Bandyopadhyay,
  Basappa, and Sood}}]{Bandyopadhyay:00}
\bibinfo{author}{\bibfnamefont{R.}~\bibnamefont{Bandyopadhyay}},
  \bibinfo{author}{\bibfnamefont{G.}~\bibnamefont{Basappa}}, \bibnamefont{and}
  \bibinfo{author}{\bibfnamefont{A.~K.} \bibnamefont{Sood}},
  \bibinfo{journal}{Phys. Rev. Lett.} \textbf{\bibinfo{volume}{84}},
  \bibinfo{pages}{2022} (\bibinfo{year}{2000}).

\bibitem[{\citenamefont{Hu et~al.}(1998)\citenamefont{Hu, Boltenhagen, and
  Pine}}]{Hu:98}
\bibinfo{author}{\bibfnamefont{Y.~T.} \bibnamefont{Hu}},
  \bibinfo{author}{\bibfnamefont{P.}~\bibnamefont{Boltenhagen}},
  \bibnamefont{and} \bibinfo{author}{\bibfnamefont{D.~J.} \bibnamefont{Pine}},
  \bibinfo{journal}{J. Rheol.} \textbf{\bibinfo{volume}{42}},
  \bibinfo{pages}{1185} (\bibinfo{year}{1998}).

\bibitem[{\citenamefont{Meyer et~al.}(1999)\citenamefont{Meyer, Asnacios,
  Bourgaux, and Kleman}}]{Meyer:99}
\bibinfo{author}{\bibfnamefont{C.}~\bibnamefont{Meyer}},
  \bibinfo{author}{\bibfnamefont{S.}~\bibnamefont{Asnacios}},
  \bibinfo{author}{\bibfnamefont{C.}~\bibnamefont{Bourgaux}}, \bibnamefont{and}
  \bibinfo{author}{\bibfnamefont{M.}~\bibnamefont{Kleman}},
  \bibinfo{journal}{Mol. Cryst. Liq. Cryst.} \textbf{\bibinfo{volume}{332}},
  \bibinfo{pages}{531} (\bibinfo{year}{1999}).

\bibitem[{\citenamefont{Grosso et~al.}(2001)\citenamefont{Grosso, Keunings,
  Crescitelli, and Maffettone}}]{Grosso:01}
\bibinfo{author}{\bibfnamefont{M.}~\bibnamefont{Grosso}},
  \bibinfo{author}{\bibfnamefont{R.}~\bibnamefont{Keunings}},
  \bibinfo{author}{\bibfnamefont{S.}~\bibnamefont{Crescitelli}},
  \bibnamefont{and} \bibinfo{author}{\bibfnamefont{P.~L.}
  \bibnamefont{Maffettone}}, \bibinfo{journal}{Phys. Rev. Lett.}
  \textbf{\bibinfo{volume}{86}}, \bibinfo{pages}{3184} (\bibinfo{year}{2001}).

\bibitem[{\citenamefont{Head et~al.}(2001)\citenamefont{Head, Ajdari, and
  Cates}}]{Head:01}
\bibinfo{author}{\bibfnamefont{D.~A.} \bibnamefont{Head}},
  \bibinfo{author}{\bibfnamefont{A.}~\bibnamefont{Ajdari}}, \bibnamefont{and}
  \bibinfo{author}{\bibfnamefont{M.~E.} \bibnamefont{Cates}},
  \bibinfo{journal}{Phys. Rev. E} \textbf{\bibinfo{volume}{64}},
  \bibinfo{pages}{061509} (\bibinfo{year}{2001}).

\bibitem[{\citenamefont{Cates et~al.}(2002)\citenamefont{Cates, Head, and
  Ajdari}}]{Cates:02}
\bibinfo{author}{\bibfnamefont{M.~E.} \bibnamefont{Cates}},
  \bibinfo{author}{\bibfnamefont{D.~A.} \bibnamefont{Head}}, \bibnamefont{and}
  \bibinfo{author}{\bibfnamefont{A.}~\bibnamefont{Ajdari}}
  (\bibinfo{year}{2002}), \bibinfo{note}{to be published, preprint cond-mat/0204162}.

\bibitem[{\citenamefont{Sierro and Roux}(1997)}]{Sierro:97}
\bibinfo{author}{\bibfnamefont{P.}~\bibnamefont{Sierro}} \bibnamefont{and}
  \bibinfo{author}{\bibfnamefont{D.}~\bibnamefont{Roux}},
  \bibinfo{journal}{Phys. Rev. Lett.} \textbf{\bibinfo{volume}{78}},
  \bibinfo{pages}{1496} (\bibinfo{year}{1997}).

\bibitem[{\citenamefont{Leng et~al.}(2001)\citenamefont{Leng, Nallet, and
  Roux}}]{Leng:01}
\bibinfo{author}{\bibfnamefont{J.}~\bibnamefont{Leng}},
  \bibinfo{author}{\bibfnamefont{F.}~\bibnamefont{Nallet}}, \bibnamefont{and}
  \bibinfo{author}{\bibfnamefont{D.}~\bibnamefont{Roux}},
  \bibinfo{journal}{Eur. Phys. J. E.} \textbf{\bibinfo{volume}{4}},
  \bibinfo{pages}{77} (\bibinfo{year}{2001}).

\bibitem[{\citenamefont{Roux et~al.}(1993)\citenamefont{Roux, Nallet, and
  Diat}}]{Roux:93}
\bibinfo{author}{\bibfnamefont{D.}~\bibnamefont{Roux}},
  \bibinfo{author}{\bibfnamefont{F.}~\bibnamefont{Nallet}}, \bibnamefont{and}
  \bibinfo{author}{\bibfnamefont{O.}~\bibnamefont{Diat}},
  \bibinfo{journal}{Europhys. lett.} \textbf{\bibinfo{volume}{24}},
  \bibinfo{pages}{53} (\bibinfo{year}{1993}).

\bibitem[{\citenamefont{Herve et~al.}(1993)\citenamefont{Herve, Roux, Bellocq,
  Nallet, and Gulik-Krzywicki}}]{Herve:93}
\bibinfo{author}{\bibfnamefont{P.}~\bibnamefont{Herve}},
  \bibinfo{author}{\bibfnamefont{D.}~\bibnamefont{Roux}},
  \bibinfo{author}{\bibfnamefont{A.-M.} \bibnamefont{Bellocq}},
  \bibinfo{author}{\bibfnamefont{F.}~\bibnamefont{Nallet}}, \bibnamefont{and}
  \bibinfo{author}{\bibfnamefont{T.}~\bibnamefont{Gulik-Krzywicki}},
  \bibinfo{journal}{J. Phys. II France} \textbf{\bibinfo{volume}{3}},
  \bibinfo{pages}{1255} (\bibinfo{year}{1993}).

\bibitem[{\citenamefont{Helfrich}(1978)}]{Helfrich:78}
\bibinfo{author}{\bibfnamefont{W.}~\bibnamefont{Helfrich}},
  \bibinfo{journal}{Z. Naturforsh} \textbf{\bibinfo{volume}{33a}},
  \bibinfo{pages}{305} (\bibinfo{year}{1978}).

\bibitem[{\citenamefont{Porte et~al.}(1991)\citenamefont{Porte, Appell,
  Bassereau, Marignan, Skouri, Billard, and Delasanti}}]{Porte:91}
\bibinfo{author}{\bibfnamefont{G.}~\bibnamefont{Porte}},
  \bibinfo{author}{\bibfnamefont{J.}~\bibnamefont{Appell}},
  \bibinfo{author}{\bibfnamefont{P.}~\bibnamefont{Bassereau}},
  \bibinfo{author}{\bibfnamefont{J.}~\bibnamefont{Marignan}},
  \bibinfo{author}{\bibfnamefont{M.}~\bibnamefont{Skouri}},
  \bibinfo{author}{\bibfnamefont{I.}~\bibnamefont{Billard}}, \bibnamefont{and}
  \bibinfo{author}{\bibfnamefont{M.}~\bibnamefont{Delasanti}},
  \bibinfo{journal}{Physica A} \textbf{\bibinfo{volume}{176}},
  \bibinfo{pages}{168} (\bibinfo{year}{1991}).

\bibitem[{\citenamefont{Roux et~al.}(1992)\citenamefont{Roux, Coulon, and
  Cates}}]{Roux:92}
\bibinfo{author}{\bibfnamefont{D.}~\bibnamefont{Roux}},
  \bibinfo{author}{\bibfnamefont{C.}~\bibnamefont{Coulon}}, \bibnamefont{and}
  \bibinfo{author}{\bibfnamefont{M.~E.} \bibnamefont{Cates}},
  \bibinfo{journal}{J. Phys. Chem.} \textbf{\bibinfo{volume}{96}},
  \bibinfo{pages}{4174} (\bibinfo{year}{1992}).

\bibitem[{\citenamefont{Diat et~al.}(1995)\citenamefont{Diat, Roux, and
  Nallet}}]{Diat:95}
\bibinfo{author}{\bibfnamefont{O.}~\bibnamefont{Diat}},
  \bibinfo{author}{\bibfnamefont{D.}~\bibnamefont{Roux}}, \bibnamefont{and}
  \bibinfo{author}{\bibfnamefont{F.}~\bibnamefont{Nallet}},
  \bibinfo{journal}{Phys. Rev. E} \textbf{\bibinfo{volume}{51}},
  \bibinfo{pages}{3296} (\bibinfo{year}{1995}).

\bibitem[{\citenamefont{Ackerson and Clark}(1984)}]{Ackerson:84}
\bibinfo{author}{\bibfnamefont{B.~J.} \bibnamefont{Ackerson}} \bibnamefont{and}
  \bibinfo{author}{\bibfnamefont{N.~A.} \bibnamefont{Clark}},
  \bibinfo{journal}{Phys. Rev. A} \textbf{\bibinfo{volume}{30}},
  \bibinfo{pages}{906} (\bibinfo{year}{1984}).

\bibitem[{\citenamefont{Richetti et~al.}(1987)\citenamefont{Richetti, Roux,
  Argoul, and Arn\'{e}odo}}]{Richetti:87}
\bibinfo{author}{\bibfnamefont{P.}~\bibnamefont{Richetti}},
  \bibinfo{author}{\bibfnamefont{J.-C.} \bibnamefont{Roux}},
  \bibinfo{author}{\bibfnamefont{F.}~\bibnamefont{Argoul}}, \bibnamefont{and}
  \bibinfo{author}{\bibfnamefont{A.}~\bibnamefont{Arn\'{e}odo}},
  \bibinfo{journal}{J. Chem. Phys.} \textbf{\bibinfo{volume}{86}},
  \bibinfo{pages}{3339} (\bibinfo{year}{1987}).

\bibitem[{\citenamefont{Argoul et~al.}(1987)\citenamefont{Argoul, Arn\'{e}odo,
  Richetti, and Roux}}]{Argoul:87}
\bibinfo{author}{\bibfnamefont{F.}~\bibnamefont{Argoul}},
  \bibinfo{author}{\bibfnamefont{A.}~\bibnamefont{Arn\'{e}odo}},
  \bibinfo{author}{\bibfnamefont{P.}~\bibnamefont{Richetti}}, \bibnamefont{and}
  \bibinfo{author}{\bibfnamefont{J.-C.} \bibnamefont{Roux}},
  \bibinfo{journal}{J. Chem. Phys.} \textbf{\bibinfo{volume}{86}},
  \bibinfo{pages}{3325} (\bibinfo{year}{1987}).

\bibitem[{\citenamefont{Grand et~al.}(1997)\citenamefont{Grand, Arrault, and
  Cates}}]{Grand:97}
\bibinfo{author}{\bibfnamefont{C.}~\bibnamefont{Grand}},
  \bibinfo{author}{\bibfnamefont{J.}~\bibnamefont{Arrault}}, \bibnamefont{and}
  \bibinfo{author}{\bibfnamefont{M.~E.} \bibnamefont{Cates}},
  \bibinfo{journal}{J. Phys. II France} \textbf{\bibinfo{volume}{7}},
  \bibinfo{pages}{1071} (\bibinfo{year}{1997}).

\bibitem[{\citenamefont{Fischer and Callaghan}(2001)}]{Fisher:01}
\bibinfo{author}{\bibfnamefont{E.}~\bibnamefont{Fischer}} \bibnamefont{and}
  \bibinfo{author}{\bibfnamefont{P.~T.} \bibnamefont{Callaghan}},
  \bibinfo{journal}{Phys. Rev. E} \textbf{\bibinfo{volume}{64}},
  \bibinfo{pages}{011501} (\bibinfo{year}{2001}).

\bibitem[{\citenamefont{Goveas and Olmsted}()}]{Goveas:01}
\bibinfo{author}{\bibfnamefont{J.~L.} \bibnamefont{Goveas}} \bibnamefont{and}
  \bibinfo{author}{\bibfnamefont{P.~D.} \bibnamefont{Olmsted}},
  \bibinfo{note}{to be published, preprint cond-mat/0104191}.

\bibitem[{\citenamefont{Spenley et~al.}(1993)\citenamefont{Spenley, Cates, and
  McLeish}}]{Spenley:93}
\bibinfo{author}{\bibfnamefont{A.}~\bibnamefont{Spenley}},
  \bibinfo{author}{\bibfnamefont{M.~E.} \bibnamefont{Cates}}, \bibnamefont{and}
  \bibinfo{author}{\bibfnamefont{T.~C.~B.} \bibnamefont{McLeish}},
  \bibinfo{journal}{Phys. Rev. Lett.} \textbf{\bibinfo{volume}{71}},
  \bibinfo{pages}{939} (\bibinfo{year}{1993}).

\bibitem[{\citenamefont{Gilmore}(1998)}]{Gilmore:98}
\bibinfo{author}{\bibfnamefont{R.}~\bibnamefont{Gilmore}},
  \bibinfo{journal}{Rev. Mod. Phys.} \textbf{\bibinfo{volume}{70}},
  \bibinfo{pages}{1455} (\bibinfo{year}{1998}).

\bibitem[{\citenamefont{Lorenz}(1963)}]{Lorenz:63}
\bibinfo{author}{\bibfnamefont{E.~N.} \bibnamefont{Lorenz}},
  \bibinfo{journal}{J. Atmos. Sci.} \textbf{\bibinfo{volume}{20}},
  \bibinfo{pages}{130} (\bibinfo{year}{1963}).

\bibitem[{\citenamefont{Guckenheimer and Holmes}(1984)}]{Guckenheimer}
\bibinfo{author}{\bibfnamefont{J.}~\bibnamefont{Guckenheimer}}
  \bibnamefont{and} \bibinfo{author}{\bibfnamefont{P.}~\bibnamefont{Holmes}},
  \emph{\bibinfo{title}{Nonlinear Oscillations, Dynamical Systems, and
  Bifurcations of Vector Fields}} (\bibinfo{publisher}{Springer, Berlin},
  \bibinfo{year}{1984}).

\bibitem[{\citenamefont{Crawford}(1991)}]{Crawford:91}
\bibinfo{author}{\bibfnamefont{J.~D.} \bibnamefont{Crawford}},
  \bibinfo{journal}{Rev. Mod. Phys.} \textbf{\bibinfo{volume}{63}},
  \bibinfo{pages}{991} (\bibinfo{year}{1991}).

\bibitem[{\citenamefont{Gang et~al.}(1993)\citenamefont{Gang, Ditzinger, Ning,
  and Haken}}]{Gang:93}
\bibinfo{author}{\bibfnamefont{H.}~\bibnamefont{Gang}},
  \bibinfo{author}{\bibfnamefont{T.}~\bibnamefont{Ditzinger}},
  \bibinfo{author}{\bibfnamefont{C.~Z.} \bibnamefont{Ning}}, \bibnamefont{and}
  \bibinfo{author}{\bibfnamefont{H.}~\bibnamefont{Haken}},
  \bibinfo{journal}{Phys. Rev. Lett.} \textbf{\bibinfo{volume}{71}},
  \bibinfo{pages}{807} (\bibinfo{year}{1993}).



\bibitem[{\citenamefont{Panizza et~al.}(1998)\citenamefont{Panizza,
  Colin, Coulon and Roux}}]{Panizza:98}
\bibinfo{author}{\bibfnamefont{P.}~\bibnamefont{Panizza}},
\bibinfo{author}{\bibfnamefont{A.}~\bibnamefont{Colin}},
\bibinfo{author}{\bibfnamefont{C}~\bibnamefont{Coulon}},  
\bibinfo{author}{\bibfnamefont{D.}~\bibnamefont{Roux}},
  \bibinfo{journal}{Eur. Phys. J. B.  } \textbf{\bibinfo{volume}{4}},
  \bibinfo{pages}{65} (\bibinfo{year}{1996}).







\end{thebibliography}
\end{document}